\documentclass[fleqn,usenatbib]{mnras}

\usepackage[T1]{fontenc}

\DeclareRobustCommand{\VAN}[3]{#2}
\let\VANthebibliography\thebibliography
\def\thebibliography{\DeclareRobustCommand{\VAN}[3]{##3}\VANthebibliography}

\newcommand{\lya}{Ly$\alpha$}
\newcommand{\zs}{Z$_\ast$}

\newcommand{\fesc}{$f_{\rm esc, LyC}$}
\newcommand{\fescrel}{$f_{\rm esc, LyC}^{\rm rel}$}
\newcommand{\ovi}{\ion{O}{vi}}

\newcommand{\muv}{$M_{\rm UV}$}
\newcommand{\muvobs}{$M_{\rm UV}^{\rm obs}$}

\newcommand{\betaint}{$\beta_{\rm int}^{1550}$}
\newcommand{\betaobs}{$\beta_{\rm obs}^{1550}$}
\newcommand{\ms}{M$_\ast$}
\newcommand{\nion}{$\dot{n}_{\rm ion}$}
\newcommand{\xiion}{$\xi_{\rm ion}$}

\newcommand{\ebv}{E(B-V)}

\newcommand{\bpass}{\textsc{bpass}}
\newcommand{\starburst}{\textsc{starburst99}}

\usepackage{graphicx}	
\usepackage{amsmath}	
\usepackage{amssymb}	
\usepackage{mathptmx}
\usepackage{txfonts}
\usepackage{subcaption}

\title[The Continuum Slope and LyC Escape at High-redshift]{The Far-Ultraviolet Continuum Slope as a Lyman Continuum Escape Estimator at High-redshift}

\author[Chisholm et al.]{
J. Chisholm,$^{1}$\thanks{E-mail: chisholm@austin.utexas.edu}  A. Saldana-Lopez,$^{2}$ S. Flury,$^{3}$ D. Schaerer,$^{2,4}$ A. Jaskot,$^{5}$ R. Amor\'{i}n,$^{6,7}$ H. Atek,$^{8}$ \newauthor  S. L. Finkelstein,$^{1}$ B. Fleming,$^{9}$ H. Ferguson,$^{10}$ V. Fern\'{a}ndez,$^{6}$ M. Giavalisco,$^{3}$ M. Hayes,$^{11}$ T. Heckman,$^{12}$ \newauthor A. Henry,$^{10}$  Z. Ji,$^{3}$ R. Marques-Chaves,$^{2}$ V. Mauerhofer,$^{13}$ S. McCandliss,$^{12}$ M. S. Oey,$^{14}$ G. Ostlin,$^{11}$  \newauthor M. Rutkowski,$^{15}$ C. Scarlata,$^{16}$ T. Thuan,$^{17}$  M. Trebitsch,$^{13}$ B. Wang,$^{18}$ G. Worseck,$^{19}$ and X. Xu$^{12}$
\\
$^{1}$Department of Astronomy, The University of Texas at Austin, 2515 Speedway, Stop C1400, Austin, TX 78712-1205, USA\\
$^{2}$Department of Astronomy, University of Geneva, 51 Chemin Pegasi, 1290 Versoix, Switzerland\\
$^{3}$ Department of Astronomy, University of Massachusetts Amherst, Amherst, MA 01002, United States\\
$^{4}$CNRS, IRAP, 14 Avenue E. Belin, 31400 Toulouse, France\\
$^{5}$ Department of Astronomy, Williams College, Williamstown, MA 01267, United States\\
$^{6}$Departamento de Astronom\'{i}a, Universidad de La Serena, Av. Juan Cisternas 1200 Norte, La Serena, Chile\\
$^{7}$ Instituto de Investigaci\'{o}n Multidisciplinar en Ciencia y Tecnolog\'{i}a, Universidad de La Serena, Ra{u}l Bitr{a}n 1305, La Serena, Chile\\
$^{8}$ Institut d’astrophysique de Paris, CNRS UMR7095, Sorbonne Universit\'{e}, 98bis Boulevard Arago, F-75014 Paris,
France\\
$^{9}$ Laboratory for Atmospheric and Space Physics, Boulder, Colorado, United States\\
$^{10}$Space Telescope Science Institute, 3700 San Martin Drive Baltimore, MD 21218, United States\\
$^{11}$The Oskar Klein Centre, Department of Astronomy, Stockholm University, AlbaNova, SE-10691 Stockholm, Sweden\\
$^{12}$ Department of Physics and Astronomy, Johns Hopkins University, Baltimore, MD 21218, United States\\
$^{13}$Kapteyn Astronomical Institute, University of Groningen, P.O. Box 800, 9700 AV Groningen, The Netherlands\\
$^{14}$Department of Astronomy, University of Michigan, Ann Arbor, MI 48109, United States\\
$^{15}$ Department of Physics and Astronomy, Minnesota State University, Mankato, MN, 56001, United States\\
$^{16}$ Minnesota Institute for Astrophysics, School of Physics and Astronomy, University of Minnesota, 316 Church Street SE, Minneapolis, MN 55455, United States\\
$^{17}$Astronomy Department, University of Virginia, Charlottesville, VA 22904, United States\\
$^{18}$ Department of Astronomy \& Astrophysics, The Pennsylvania State University, University Park, PA 16802, USA\\
$^{19}$ Institut f\"{u}r Physik und Astronomie, Universit\"{a}t Potsdam, Karl-Liebknecht-Str. 24/25, D-14476 Potsdam, Germany
}

\date{Accepted XXX. Received YYY; in original form ZZZ}

\pubyear{2022}

\begin{document}
\label{firstpage}
\pagerange{\pageref{firstpage}--\pageref{lastpage}}
\maketitle

\begin{abstract}
    Most of the hydrogen in the intergalactic medium (IGM) was rapidly ionized at high-redshifts. While observations have established that reionization occurred, observational constraints on the emissivity of ionizing photons at high-redshift remains elusive. Here, we present a new analysis of the Low-redshift Lyman Continuum Survey (LzLCS) and archival observations, a combined sample of 89 star-forming galaxies at redshifts near 0.3 with \textit{Hubble} Space Telescope observations of their ionizing continua (or Lyman Continuum, LyC). We find a strong ($6\sigma$ significant) inverse correlation between the continuum slope at 1550~\AA\ (defined as F$_\lambda \propto \lambda^{\beta^{1550}_{\rm obs}}$) and both the LyC escape fraction (\fesc) and \fesc\ times the ionizing photon production efficiency ($\xi_{\rm ion}$). On average, galaxies with redder continuum slopes have smaller \fesc\ than  galaxies with bluer slopes due to the higher dust attenuation in redder galaxies. More than 5\% (20\%) of the LyC emission escapes galaxies with \betaobs~$< -2.1$ (-2.6). We find strong correlations between \betaobs\ and the gas-phase ionization ([\ion{O}{iii}]/[\ion{O}{ii}] flux ratio; at 7.5$\sigma$ significance), galaxy stellar mass (at 5.9$\sigma$),  the gas-phase metallicity (at 4.6$\sigma$), and the observed FUV absolute magnitude (at 3.4$\sigma$). Using previous observations of \betaobs\ at high-redshift, we estimate the evolution of \fesc\ with both redshift and galaxy magnitude. The LzLCS observations suggest that fainter and lower mass galaxies dominate the ionizing photon budget at higher redshift, possibly due to their rapidly evolving metal and dust content. Finally, we use our correlation between \betaobs\ and  \fesc~$\times$~$\xi_{\rm ion}$ to predict the ionizing emissivity of galaxies during the epoch of reionization. Our estimated emissivities match IGM observations, and suggest that star-forming galaxies emit sufficient LyC photons into the IGM to exceed recombinations near redshifts of 7--8. 
\end{abstract} 

\begin{keywords}
dark ages, reionization, first stars -- galaxies: starburst -- ultraviolet: galaxies
\end{keywords}



\section{Introduction}
\label{intro}
Observations of high-redshift quasars indicate that the intergalactic medium (IGM) underwent a large-scale phase change near $z\sim5-10$ \citep{becker, fan2006, banados, becker21}. At these redshifts, gas between galaxies rapidly transitions from being largely neutral to being largely ionized. This \lq{}\lq{}epoch of reionization\rq{}\rq{} seeds the subsequent large-scale structure of the Universe \citep{gnedin00}, impacts the interpretation of the cosmic microwave background radiation \citep{planck16}, and sets the IGM temperature \citep{Miralda}.

Determining the structure and evolution of reionization requires uncovering what emitted the first hydrogen ionizing, or Lyman Continuum (LyC), photons. Whether the sources of ionizing photons are broadly distributed spatially, or more concentrated impacts the morphology of reionzation \citep{Kulkarni}. The timing and duration of reionzation strongly depend upon whether massive stars, active galactic nuclei (AGN), or evolved stellar remnants are key contributors to reionization \citep{robertson13, madau14, robertson15, rosdahl18}.

The initial debate on the sources of reionization centered around whether AGN or star-forming galaxies reionized the early IGM \citep{ouchi, hopkins08, claud09, finkelstein12, robertson13, madau15}. However, current observations and simulations find too few active galactic nuclei during the epoch of reionization to dominate the ionizing photon budget \citep{fontanot, trebitsch17, ricci, matsuoka18, Kulkarni19,  shen20, claude20, dayal20, yung21, trebitsch21, Jiang22}. The observational landscape of high-redshift AGN may dramatically change as the sensitivity of the JWST opens a new window onto the epoch of reionization, but the current observational picture suggests that star-forming galaxies reionized the IGM. Coupled with the recent success of LyC observations from local galaxies (see below), the current debate has shifted to whether bright massive galaxies that are clustered in over densities or widely-distributed low-mass galaxies emitted the requisite ionizing photons \citep{finkelstein19, naidu19, mason19, finkelstein21, naidu21, matthee21}.

Determining the ionizing emissivity ($\dot{n}_{\rm ion}$) of a given population (e.g. AGN or star-forming) at high redshift requires estimating the number density of sources and the number of ionizing photons that each source emits into the IGM. This can be cast in terms of the UV luminosity density ($\rho_{\rm UV}$), the production efficiency of ionizing photons ($\xi_{\rm ion}$, or the number of ionizing photons per FUV luminosity), and the fraction of LyC photons that escape each source (\fesc). Numerically, \nion\ is the product of these three quantities as
\begin{equation}
    \dot{n}_{\rm ion} = \rho_{\rm UV}\times\xi_{\rm ion}\times f_{\rm esc, LyC} \label{eq:nion} .
\end{equation}
All of these quantities likely vary with redshift and with galaxy property (e.g. stellar mass or far-ultraviolet absolute magnitude), but each quantity can in principle be estimated from galaxies in the epoch of reionization.

Deep Hubble Space Telescope observations have estimated $\rho_{\rm UV}$ out to redshifts as high as 10 \citep{bouwens15, finkelstein15, mcleod, livermore17, Mehta, atek18, oesch18, yue, bouwens21, bouwens22}. Similarly,  deep rest-frame far-ultraviolet (FUV) spectroscopy \citep{stark15, mainali18, mengtao, hutchison}  and \textit{Spitzer} rest-frame optical \citep{bouwens16, debarros, endsley21} observations of epoch of reionization-era systems find hard radiation fields indicative of high \xiion\ ($\sim10^{25.3-26}$~[s$^{-1}$/erg~s$^{-1}$~Hz$^{-1}$]). Combined, these observations hint that star-forming galaxies can reionize the IGM if \fesc~$> 5-20$\% \citep{ouchi, robertson13, robertson15, finkelstein19, mason19, naidu19, naidu21, matthee21}.

The neutral IGM precludes  direct LyC detections at $z > 4$ in a statistical sense \citep{inoue14, becker21}, outside of rare ionized regions \citep{endsley21, endsley22}. However, \fesc\ can be directly measured at $z<4$ \citep{bergvall, leitet11, borthakur, izotov16b, izotov16a, vanzella16, leitherer16, shapley16, izotov18a, izotov18b, steidel18, fletcher19, wang19, rivera19, ji, izotov21, davis, marques-chaves, xu22}.  To overcome the neutral high-redshift IGM, recent work has correlated indirect observables with \fesc\ at these low redshifts. Tentative correlations, with appreciable scatter and significant non-detections \citep{rutkowski17}, have been found between the Ly$\alpha$ emission properties \citep{verhamme15, rivera-thorsen17, izotov18b, steidel18, gazagnes20, izotov21}, ISM absorption properties \citep{reddy16, chisholm17, gazanges, chisholm18, steidel18}, \ion{Mg}{ii} resonant emission lines \citep{henry18, chisholm20, witstok, xu22}, and optical emission line ratios \citep{nakajima14, jaskot, wang19}. Recent  simulations have tested these indirect indicators and have found modest efficacy with strong impacts from geometry and dust \citep{mauerhofer, katz22}. These first glimpses of small samples of LyC emitters at lower redshifts provide a blueprint to estimate \fesc\ during the epoch of reionization.

While there are promising LyC tracers, LyC emitting galaxies have been elusive and the sample sizes of confirmed LyC detections was still modest. The small LyC samples may obscure trends or introduce false trends in LyC escape detections. The Low-Redshift Lyman Continuum Survey \citep[LzLCS;][]{flurya} is a 134 orbit Hubble Space Telescope project that used the Cosmic Origins Spectrograph (COS) to target 66 new $z\sim 0.3$ LyC emitting candidates. The LzLCS was developed to explore the  galaxy parameter range relevant to cosmic reionization to assess the validity of a variety of indirect LyC tracers \citep{flurya}. The LzLCS sample has a dynamic range in three purported LyC diagnostics: H$\beta$ equivalent width, the [\ion{O}{iii}]/[\ion{O}{ii}] flux ratio, and the slope of the FUV stellar continuum ($\beta$). This range of parameters stringently tests indirect trends of LyC escape, establishing whether they scale with \fesc\ at low-redshifts \citep{fluryb}.

The LzLCS is a resounding success. Combined with archival HST/COS observations, the full LzLCS plus archival survey contains 89 LyC observations. More than half, 55\%, of these galaxies have LyC emission detected above the background at the  $>$97.7\% confidence level and the other 45\% have robust upper limits \citep{flurya}. This provides a definitive local sample to explore indirect diagnostics in \fesc. \citet{fluryb} found that many of the classical LyC diagnostics have appreciable scatter, but in general more highly ionized, vigorously star-forming, and compact galaxies are more likely to emit ionizing photons. The Ly$\alpha$ equivalent width, escape fraction, and velocity separation strongly correlate with \fesc\ \citep{fluryb}. LyC emitters within the LzLCS are deficient in low-ionization optical metal emission lines, specifically [\ion{S}{ii}], indicating that LyC emitters are  relatively deficient in low-ionization gas  \citep{wang21}. The COS FUV spectral range also provides invaluable probes of the \ion{H}{i} in the galaxies with the Lyman Series absorption lines. These \ion{H}{i} lines suggest that LyC emission escapes through regions in the galaxy with low neutral gas covering fraction and low dust attenuation \citep{Saldana}. As such, \fesc\ can be robustly predicted using the combination of \ion{H}{i} absorption and dust attenuation.

Here we extend the previous LzLCS work to a detailed analysis of the FUV spectral slope, \betaobs, derived from stellar population fits and their relation to the \fesc. \betaobs\ traces both the stellar population parameters and the dust attenuation in the LzLCS galaxies, and has been theoretically predicted to correlate with \fesc\ \citep[e.g.,][]{zackrisson13}. In \autoref{obs} we describe the LzLCS sample and observations.  \autoref{cont_fit} describes the modeling of the FUV stellar continua, and the derivation of stellar continuum properties. We explore correlations with \betaobs\ (\autoref{beta}) before introducing the strong relation between \betaobs\ and the \fesc\ (\autoref{fesc}). The physical picture presented by the \betaobs-\fesc\ relation is explored in \autoref{escape}. We conclude in \autoref{reionization} by using the \betaobs-\fesc\ relation to make predictions for \fesc\  and \nion\ during the epoch of reionzation to predict which galaxies heavily contribute to reionization. In \autoref{ebv} we also detail the impact of different assumed extinction laws.  Throughout this paper we use AB magnitudes; a standard cosmology with $H_0$~=~70~km~s$^{-1}$, $\Omega_m$~=~0.3, and $\Omega_\Lambda$~=~0.7; and stellar metallicities that are relative to a solar value of 0.02.

\section{Observations}
\label{obs}

Here we describe the observations of the local LyC emitting galaxies. We use three subsets of data: the Low-redshift Lyman Continuum Survey (\autoref{lzlcs}), archival observations of 23 LyC emitters \citep[\autoref{izotov}; ][]{izotov16a, izotov16b, izotov18a, izotov18b, wang19, izotov21}, and finally a coaddition of all the LzLCS plus archival spectra. We combine the LzLCS and archival samples into one cohesive sample that we refer to as the full LzLCS sample. 

\subsection{The LzLCS} 
\label{lzlcs}
We use data from the Low-redshift Lyman Continuum Survey \citep[LzLCS; PI: Jaskot, HST Project ID: 15626; ][]{flurya}. The LzLCS is a large Hubble Space Telescope program consisting of 134 orbits targeting 66 star-forming galaxies at redshifts between 0.219--0.431. We chose these redshifts to shift the LyC onto the sensitive spectral portion of the COS detector. These 66 galaxies were selected using optical observations from the 16th Data Release of the Sloan Digital Sky Survey \citep[SDSS; ][]{Ahumada}, and non-ionizing FUV observations from the {\it GALaxy Evolution eXplorer} \citep[GALEX; ][]{galex}. The LzLCS was designed to test the efficacy of three potential indicators of LyC escape by creating three separate, but overlapping, samples with a large dynamic range of star formation rate surface density (as probed by the $\Sigma_{\rm SFR}$ measured from H$\beta$ emission), nebular ionization state (as probed by the [\ion{O}{iii}]/[\ion{O}{ii}] optical emission line ratio), and UV continuum slopes (as measured by the {\it GALEX} FUV-NUV slope). The resultant sample probes the full property range expected at high-redshift  to test the impact of various properties on \fesc. 

The details of the data reduction are given in \citet{flurya}, but here we summarize the procedure. Each galaxy was observed for between 1--5~orbits with the low-resolution G140L grating of the Cosmic Origins Spectrograph \citep[COS; ][]{cos}. We used the G140L/800 mode \citep{redwine}, which has observed wavelength coverage between 800-1950\AA\ at a spectral resolution of $R \sim 1000$ at 1100\AA. This extremely blue-optimized COS configuration places the LyC onto the sensitive portion of the COS detector and affords rest-frame coverage of 600-1480\AA\ for the median redshift of the LzLCS. In addition to the crucial LyC, the LzLCS observations sample the very blue of the non-ionizing FUV continuum. 

The COS observations were flat-fielded, wavelength calibrated, and flux calibrated using CALCOS v3.3.9. The individual exposures were co-added in the raw counts using \textsc{faintcos} \citep{Makan21}. \textsc{faintcos} directly uses many time-baselines of archival COS observations to re-estimate the dark current,  uses the scattered light model from \citet{worseck16} to improve the flux calibration at faint flux levels, and accounts for the Poisson uncertainties of the photon-counting COS FUV detector. \citet{flurya} also details the accounting of the geocoronal emission to address possible LyC contamination. 

The individual COS spectra were corrected for the foreground Milky Way reddening using the \citet{green18} dust maps and the Milky Way extinction curve \citep{fitzpatrick99}. We then measured the observed counts in 20\AA\ regions in the rest-frame LyC. The inferred LyC counts and background levels, along with a Poisson distribution, are used in a survival function to determine the probability that the signal arises through chance realizations from the background distribution.  If the survival analysis confirms that the flux is above the background counts at the $>97.7$\% confidence level, we classify these as LyC detections, and we refer to them as galaxies with \lq{}\lq{}Detected LyC\rq{}\rq{} for the rest of the paper. We also checked the 2-dimensional spectra by eye to ensure that these detections are not spikes in background counts.  The galaxies with LyC flux  $<97.7$\% above the background are called \lq{}\lq{}non-detected LyC\rq{}\rq{} galaxies throughout this paper. These detection thresholds do not mean that the non-detected galaxies do not emit LyC photons, rather the LyC is not significantly detected given the observed backgrounds. Of the 66 galaxies in the LzLCS, 35 have LyC photons detected at $>97.7$\% confidence level. 

All LzLCS galaxies have archival SDSS optical spectroscopy \citep{Ahumada}.  These optical emission lines describe the nebular emission line properties and the gaseous conditions. In \autoref{beta}, we use these emission lines to explore the relation between nebular conditions and the FUV slope. The nebular reddening values, E(B-V)$_{\rm gas}$, are determined in an iterative way where the temperature, density, and extinction \citep[assuming the extinction law from ][]{cardelli} are all iterated until the Balmer emission line flux ratios converge \citep{flurya}. In this way, the temperature and density dependence of the Balmer line ratios are accounted for when determining the internal reddening of the galaxies. The nebular electron temperatures (T$_e$) are directly measured using the [\ion{O}{iii}]~4363~\AA\ auroral line for 54 galaxies. The extinction-corrected emission line fluxes are used to determine line-ratios of certain ions (e.g., [\ion{O}{iii}]~5007/[\ion{O}{ii}]~3726+3729~\AA) which serve as a proxy of the gas ionization.  Finally, the [\ion{S}{ii}]~6717, 6731~\AA\ \AA\ doublet is used to determine the electron density of the gas and the metallicities are measured from the inferred electron temperatures, densities, and attenuation-corrected fluxes. Metallicities of the 12 galaxies without auroral line detections were determined using the strong-line calibration from \citet{pilyugin06}.  

\subsection{Archival Data} 
\label{izotov}
In combination with the LzLCS, we also use a compilation of 23 galaxies drawn from the literature \citep{izotov16a, izotov16b, izotov18a, izotov18b, wang19, izotov21}. These galaxies are also observed with COS on HST and their reduction is handled in the same exact way as the rest of the LzLCS.  This produces a homogeneous sample of 89 galaxies with LyC observations. Of the archival sample, we detect the LyC above the background at a $>$97.7\% confidence level in 15 galaxies. This leads to a total of 50 detections in the full LzLCS plus archive sample. 

\subsection{Spectral co-additions}
\label{stack}
In order to illustrate the changes of the stellar features at high significance, we also coadd the spectra to increase the resultant signal-to-noise. The details of this process are found in Flury et al. (in preparation). We coadded all of the full LzLCS sample and made two separate composite spectra: (1) of all galaxies with detected LyC emission and (2) of all galaxies without LyC detections. We take care to mask intervening ISM, Milky Way absorption features, as well as the Ly$\alpha$ and \ion{O}{i}~1302 and 1306~\AA\ geocoronal emission lines, in each individual galaxy before co-adding the spectra. The individual spectra are then normalized to the median flux density at 1100~\AA\ and the spectra are coadded in the rest-frame. The uncertainties on the flux density are determined by bootstrapping the individual fluxes that comprise the stack. This high signal-to-noise stack is used in the next section to compare to stellar population synthesis models.

\section{Stellar Continuum Fitting}
\label{cont_fit}
\subsection{Fitting procedure}
In \citet{Saldana}, we fit the observed stellar continua of the LzLCS and archival samples, following \citet{chisholm19}, as a linear combination of multiple single age, single metallicity stellar continuum models (see \autoref{models} for a discussion of the models used). Here, we use those continuum fits and briefly describe the fitting procedure, but refer the reader to \citet{chisholm19} and \citet{Saldana} for in depth discussions on the procedure and the associated assumptions. 

We assume that the observed stellar continuum ($F(\lambda)$) is the product of the intrinsic stellar continuum ($F_i(\lambda)$) attenuated by a uniform dust screen. The treatment of the dust screen geometry impacts the interpretation of the stellar continuum fitting \citep{calzetti94, vasei, gazanges}, but it is challenging to distinguish the appropriate model with many degenerate effects \citep{calzetti2014}. Therefore, we assume a uniform dust screen. The intrinsic flux is modeled as a linear combination of a set of $j$ stellar continuum models with different ages and metallicities ($M_j(\lambda)$) times a weight ($X_j$) that describes the contribution of each $M_j$ to the total $F_i(\lambda)$. These $X_j$ can have values greater than or equal to 0.  Mathematically, this is defined as
\begin{equation}
    F(\lambda) =F_i(\lambda) \times  10^{-0.4 {\rm E(B-V)} k(\lambda)} = 10^{-0.4 {\rm E(B-V)} k(\lambda)}\Sigma_j \left(X_j \times M_j(\lambda)\right) .
    \label{eq:linear}
\end{equation}
We use the reformulated \citet{reddy16} attenuation curves that is prescribed with a dust law, k$(\lambda$), and the continuum color excess (\ebv). The \citet{reddy16} law is similar in shape to the \citet{calzetti} law, but with a lower normalization \citep[the E(B-V) is inferred to be 0.008~mag redder than the Calzetti attenuation law; see figure 5 of ][]{reddy16}. In \autoref{ebv}, we introduce analytic relations to convert between different attenuation laws. The \citet{reddy16} law is observationally defined in the observed wavelength regime down to 950\AA\, while the \citet{calzetti} is only defined to 1250\AA.  We use the python package {\sc LMFIT} \citep{lmfit} to determine the best-fit values for each of the $X_j$ linear coefficients and the \ebv\ values. 

To fit the spectra, we place the galaxies into the rest-frame using the SDSS redshifts \citep{Ahumada} and normalize the flux density to the median rest-frame flux density between 1070--1100~\AA. The rest-frame FUV contains many non-stellar features, such as foreground Milky Way absorption lines, geocoronal emission features, and ISM absorption lines from the targeted galaxies. These features do not arise from the stellar populations and should not be included in the fitting process. We mask these regions of the spectra by hand. We also mask out regions of low signal-to-noise ratios (less than one) to avoid fitting continuum noise. We tested whether including low signal-to-noise data improves the fitting and found a poorer match between models and observations when including lower quality data. We fit the rest-frame wavelengths between 950-1345~\AA\ to avoid heavy contamination of the Lyman Series lines at bluer wavelengths. The median reduced $\chi^{2}$ is 1.08 \citep{Saldana}. Detailed comparisons of models and individual spectra are given in figures 1-2 of \citet{Saldana}. 

Errors on the stellar continuum fits were calculated by modulating the observed flux density with the error spectra at each pixel, refitting the spectrum, and then tabulating the associated stellar continuum properties \citep{Saldana}. We repeated this process 500 times for each galaxy to build a sample distribution of stellar population fits. We measured the standard deviations of these Monte Carlo distributions to determine the standard deviations on all fitted parameters ($X_j$ and \ebv) and the other inferred parameters. 

\subsection{Stellar continuum models}
\label{models} 

\begin{figure}
\includegraphics[width = 0.5\textwidth]{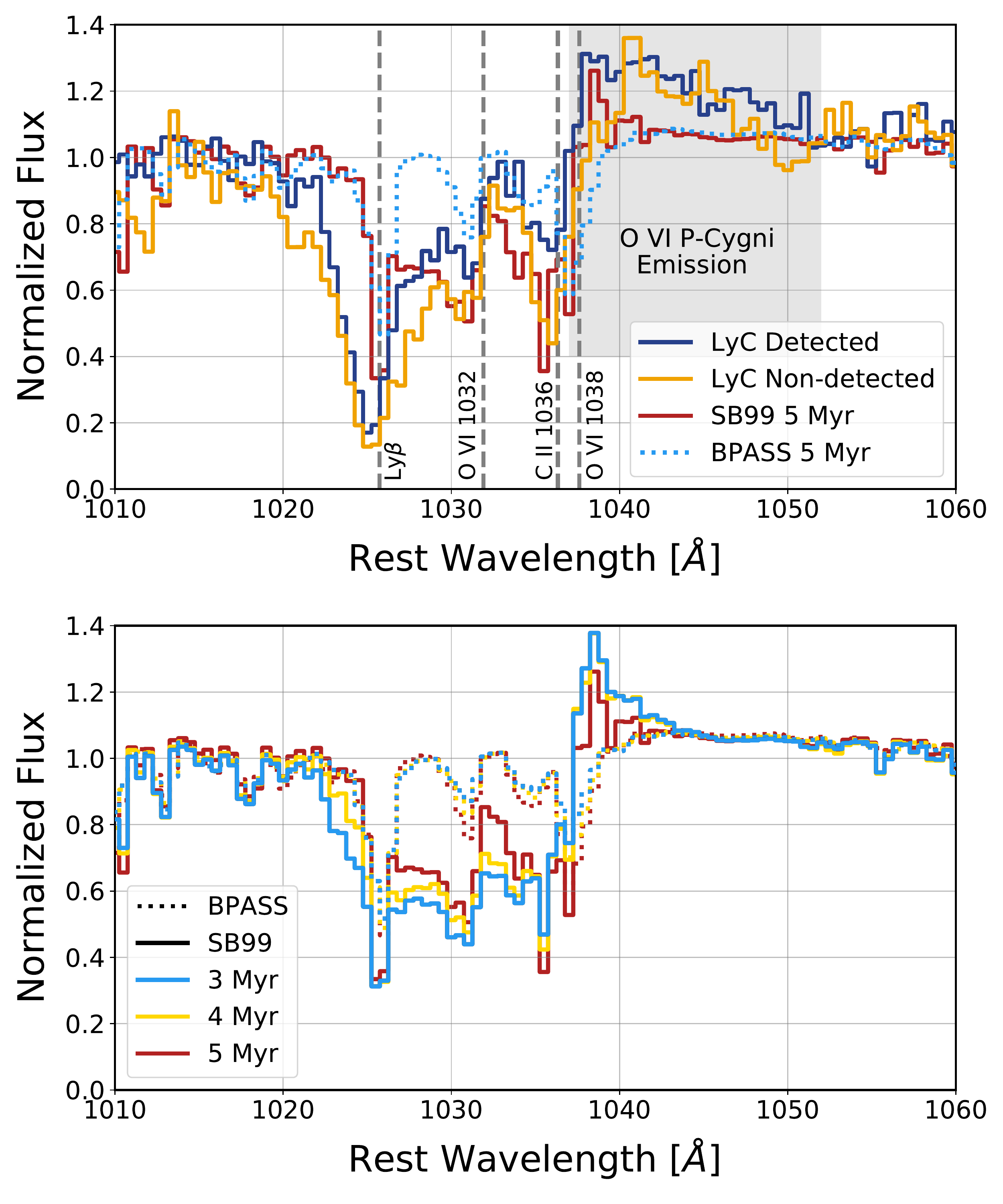}
\caption{\textbf{Upper Panel:} The \ovi~1032,1038~\AA\AA\ doublet stellar wind feature for the co-added full sample of LzLCS plus archival galaxies with LyC detections (blue line) and non-detections of the LyC (gold line). The \ovi\ profile has two main components: broad absorption that extends to 1025~\AA\ and broad emission that extends out to 1050~\AA. Strong ISM absorption lines are marked by gray vertical dashed lines. Overlaid are two fully theoretical stellar population models from \starburst\ (red line) and \bpass\ (blue dotted line), with similar parameters as the median of our fits (age of 5~Myr and metallicity of 0.2~Z$_\odot$). The \bpass\ model does not match either the observed P-Cygni absorption nor emission, while the \starburst\ model matches the absorption profile but not the full extent of the emission. \textbf{Lower Panel:} Stellar population synthesis models at a range of ages: 3 (blue), 4 (gold), and 5~Myr (red). We include both \starburst\ (solid lines) and \bpass\ (dotted lines) models. The \ovi\ profile weakens with increasing age for the \starburst\ models, indicating that \ovi\ is sensitive to the stellar population age. The \ovi\ region of the \bpass\ models hardly varies with stellar population age. The strong \ovi\ P-Cygni profile suggests that the LzLCS galaxies have very young stellar populations. }
\label{fig:ovi}
\end{figure}

To fit the observed LzLCS FUV spectra, we use fully theoretical single-age bursts of star formation to model the very young stellar continua. As in \citet{Saldana}, we include stellar models with a range of single ages of 1, 2, 3, 4, 5, 8, and 10~Myr  with metallicities in the range of 0.05, 0.2, 0.4, and 1.0 Z$_\odot$. These ages were carefully chosen to sample the spectral variations of the FUV stellar continuum \citep{chisholm19}. Each stellar model (defined by an age and metallicity) includes a  modeling of the nebular continuum by placing each single-age model into {\sc cloudy} v17.0 \citep{ferland}, assuming that all of the ionizing photons are absorbed by the nebular gas, and tabulating the resultant nebular continuum. The nebular continuum is then added back to original stellar model \citep{chisholm19}. The nebular continuum does not heavily affect the wavelengths covered by the LzLCS observations, but contributes to the continuum slope at 1550~\AA\ \citep{raiter}. We tested including older ages (up to 40~Myr) and found that they do not significantly change the inferred stellar population properties of the very young LzLCS spectra. However, the fully theoretical, high-resolution \textsc{starburst99} models do not densely sample the H-R diagram  at effective temperatures below 15,000~K \citep{claus2010}. The cooler, high-resolution models have artificially bluer continuum slopes that makes older stellar populations appear much bluer than expected for their temperatures. A similar resolution issue is found in the \bpass\ models \citep{chisholm19}. Thus, we choose the maximum  model age range of 10~Myr that includes high-resolution models that accurately probe the stellar spectral features and the continuum shape. We fully describe the observed stellar continuum using 28 free parameters (each $X_j$) for the stellar models and one for the dust (for a total of 29 free parameters).  We used the super-solar (2~Z$_\odot$) models for the three galaxies from \citet{wang19} which require higher  metallicities. 

We use the \starburst\  models, convolved to the G140L spectral resolution, as our base stellar population models \citep{claus99, claus2010}. We also fit the observed stellar  continua with \bpass\ v2.2.1 stellar continuum models that include binary evolution \citep{eldridge17, stanway18} to test the impact of the assumed stellar continuum model on the stellar population results. To do this, we create a nebular continuum for the \bpass\ models using the same method as the \starburst\ models.  

The LzLCS 1010--1060~\AA\ wavelength range contains a host of stellar spectral features that are sensitive to the stellar population properties. In particular, the top panel of \autoref{fig:ovi} compares the age sensitive \ovi\ P-Cygni stellar wind feature that is strong in both the LyC detected (blue line) and non-detected (gold line) coadded spectra. Overlaid on these coadded spectra are single age and single metallicity \starburst\ (red solid line) and \bpass\ (blue dotted lines) models. We use a 5~Myr and 0.2~Z$_\odot$ model because this is the median fitted values of the sample. The \starburst\ model roughly matches the observed broad \ovi\ absorption, while the \bpass\ model does not match the observations. The bottom panel of \autoref{fig:ovi} expands the age range of the models and shows that the absorption and emission component of \ovi\ strongly depends on the stellar population age. By the time the \starburst\ stellar population models reach 10~Myr there is very little \ovi\ emission present in either model, indicating that the average observed LzLCS FUV stellar population is dominated by $<10$~Myr stellar populations. This suggests that our choice of young stellar models is adequate to match the observed stellar continuum. 

The \starburst\ models, in general, more closely reproduce the observed \ovi\ feature which is likely due to difference in the treatment of X-rays in the stellar winds of the two models. Since the \ovi\ profile is the most obvious stellar feature in the very blue wavelength regime of the LzLCS spectra, we hereafter use the \starburst\ models as our models. However, the inferred FUV stellar continuum slopes, the key observable used here, do not significantly change when using \bpass\ models.  

\subsection{Inferred stellar population properties}

Here we describe the various properties derived from these stellar fits. These properties come in two different types: (1) directly fitted to the observations,  and (2) inferred from the stellar population fits. The directly fitted parameters are the 29 values in \autoref{eq:linear}, including the $X_j$ and the E(B-V) values. All of these parameters describe the stellar continuum in the observed FUV (near 1100~\AA) and should be considered light-weighted properties at these wavelengths. All the parameters are found in tables A1-A4 of the appendix of \citet{Saldana}.  

\subsubsection{Properties derived from fitted parameters}
From the individual fitted parameters (X$_j$ and E(B-V)) we provide light-weighted estimates of the stellar population properties. We scale the individual model parameters by the fit weights to infer the light-weighted age as
\begin{equation}
    \text{Age} = \frac{\sum_i X_j \text{Age}_j}{\sum_i X_j}
\end{equation}
and stellar metallicity (\zs) as
\begin{equation}
    Z_\ast = \frac{\sum_i X_j Z_j}{\sum_j X_j} .
\end{equation}
The median Age and $Z_\ast$ of the LzLCS sample is 4.6~Myr and 0.22~Z$_\odot$, respectively. 
\begin{figure}
\includegraphics[width = 0.5\textwidth]{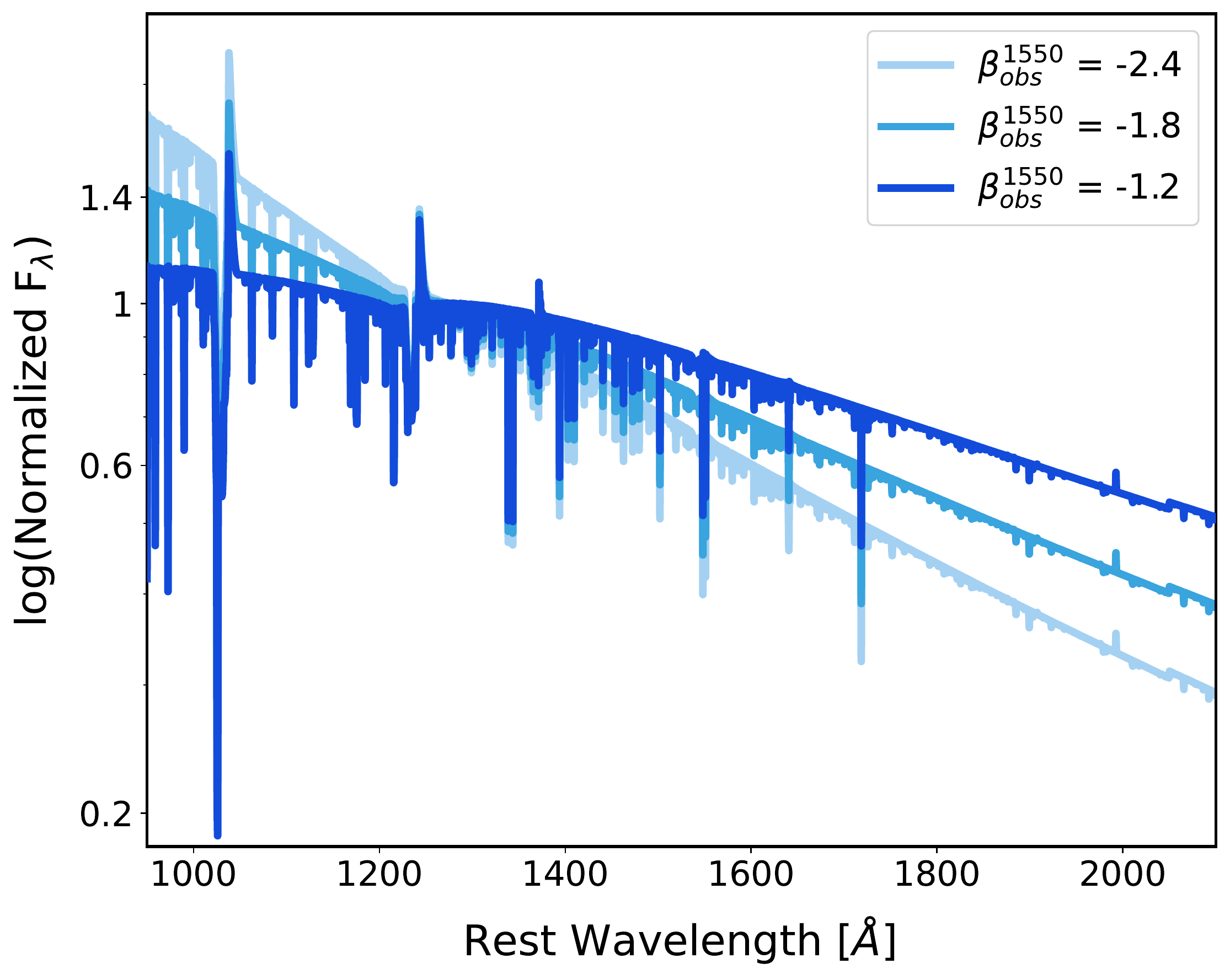}
\caption{Illustration of the impact of dust attenuation on the observed stellar continuum slope (\betaobs) keeping a constant underlying stellar population. All three spectra have the same underlying stellar population and are normalized near 1200~\AA,  but have a varying uniform dust attenuation using the \citet{reddy16} law (E(B-V) = 0.05, 0.15, 0.25 for the light blue, blue, and dark blue lines respectively). This leads to three markedly different stellar continuum slopes (\betaobs$= -2.4, -1.8, -1.2$, respectively).  }
\label{fig:beta_demo}
\end{figure}

\subsubsection{Properties derived from stellar continuum fit}\label{features}
We derive stellar continuum properties from the best-fit stellar continuum. In particular, we derive the stellar continuum slope ($\beta$), the absolute magnitude (\muvobs), \fesc, and the ionizing photon production efficiency (\xiion).

The most important property that we measure in this paper is the slope of the FUV stellar continuum at a given wavelength, $\beta^{\lambda}$. We derive the slope between rest-frame 1300~\AA\ and 1800~\AA\ (an average wavelength of 1550~\AA) by assuming that 
\begin{equation}
\label{equation:beta}
    F_{\lambda} \propto \lambda^{\beta^{1550}} .
\end{equation}
We fit for $\beta^{1550}$ by extrapolating the best-fit stellar continuum model of the LzLCS COS observations described in \autoref{cont_fit} to redder wavelengths. We then fit for \betaobs\ assuming a power-law model in  the 1300-1800~\AA\ wavelength range of the stellar continuum fit. We tested two other methods to determine $\beta^{1550}$ by using (1) several small wavelength regions in the stellar continuum models without strong stellar or ISM features and (2) fitting for the slope between two points (1300~\AA\ and 1800~\AA). We found that all three models produced consistent $\beta^{1550}$ estimates. Since we fit both the stellar population properties (giving the intrinsic continuum slope) and dust attenuation (giving the observed spectral slope), we estimate both the observed continuum slope, \betaobs, and the intrinsic slope, \betaint,  which is the continuum slope without dust attenuation. 

We use \betaobs\ from the stellar continuum modeling instead of the $\beta$ observed in the HST/COS data  \citep[$\beta^{\rm COS}$, or $\approx\beta^{\rm 1150}$; ][]{flurya} because it probes the redder wavelengths ($\sim1550$~\AA) that are observed at higher redshift \citep{finkelstein12, dunlop13, bouwens14,  wilkins16,  Bhatawdekar21, sandro21}. The \betaobs\ values scale strongly (Kendall's rank coefficient of $\tau = 0.46$ and p-value of $1\times10^{-10}$) with $\beta_{\rm COS}$ from \citet{flurya}. However, $\beta_{\rm COS}$ values are 14\% redder than \betaobs\ measured here, largely because the reddening curve is steeper at bluer wavelengths (see the discussion in \autoref{beta}) and the intrinsic stellar continuum is slightly flatter between 900-1200~\AA\ than at 1500~\AA\ \citep{claus99}. \autoref{fig:beta_demo} shows the same stellar continuum model attenuated by three different \ebv\ values. These three \ebv\ values produce three different \betaobs\ values that bracket the observed LzLCS range. \autoref{fig:beta_demo} shows unique and distinguishable $\beta^{1150}$ values for the three different \ebv, but $\beta^{1150}$ is much flatter, especially for high attenuation case, than $\beta^{1550}$. Thus, spectral slopes estimated at 1150~\AA\ are not the same as at 1550~\AA. To compare low-redshift observations to high-redshift observations, we must measure $\beta$ at similar wavelengths. 

\begin{figure}
\includegraphics[width = 0.5\textwidth]{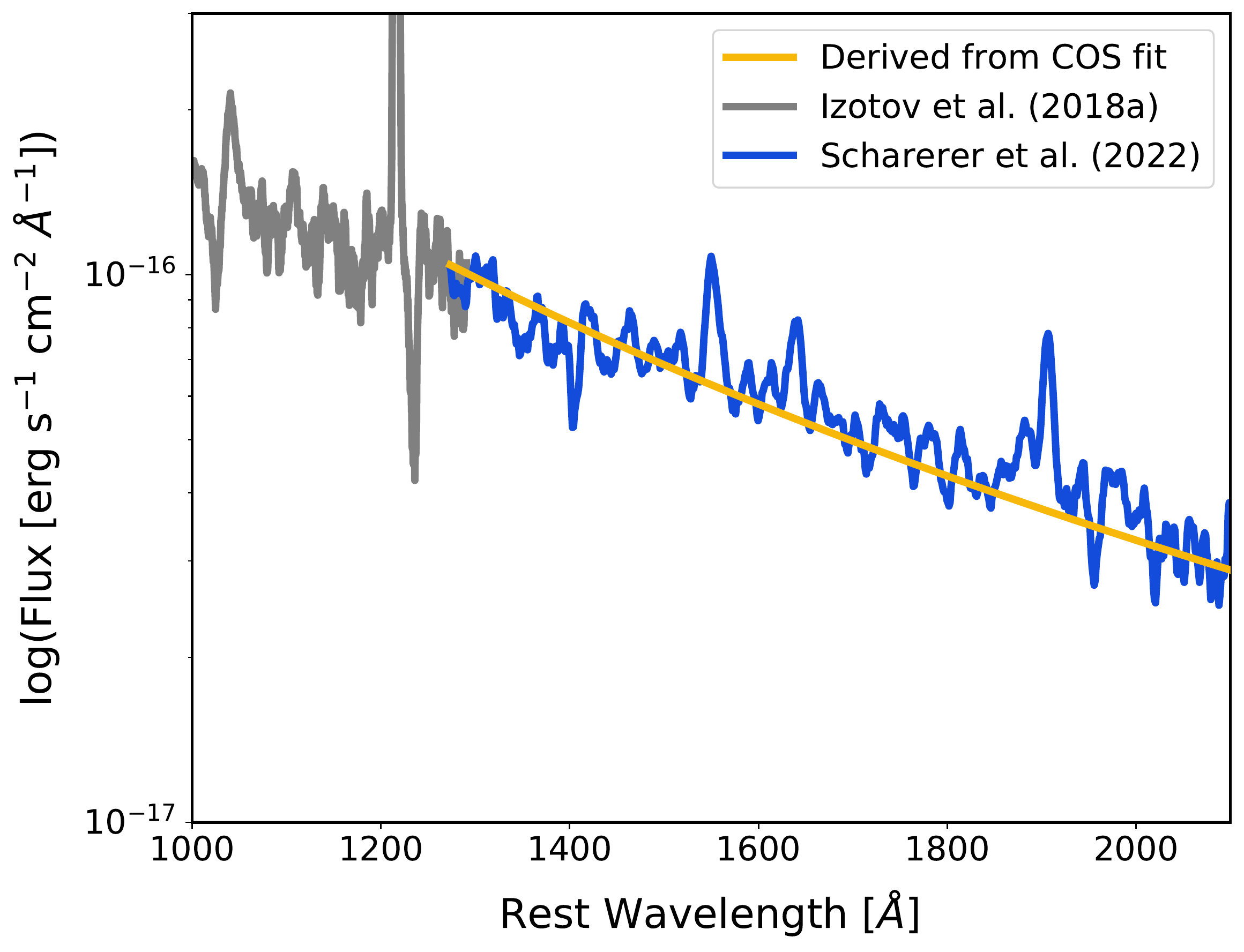}
\caption{Typical LzLCS data \citep[gray; ][]{izotov16b} only probe the blue portion of the spectrum, but recent observations from \citet{schaerer22} of J1243+4646 (blue data) compare the extrapolated slope from the LzLCS data (gold line) to the observed continuum slope at 1550~\AA. The extrapolated LzLCS continuum slope accurately matches the slope of the redder observations. }
\label{fig:beta_observations}
\end{figure}

How well do the extrapolated fits recover the actual \betaobs?  Recent HST/STIS observations by \citet{schaerer22} provide redder observations to test the extrapolated and actual \betaobs\ values. These redder observations are unavailable for the full LzLCS sample, but provide a test of our extrapolation for eight LyC emitters. \autoref{fig:beta_observations} shows the extrapolated continuum slope from the LzLCS model in gold. The gold line was estimated by fitting the stellar population model to the G140L data between rest-frame wavelengths of 950--1200~\AA\ (gray data).  The redder STIS data, shown in dark blue, has a similar slope as the LzLCS fit even though it was not included in the fit (\betaobs~=-2.56 from the models and -2.50 from the STIS observations). This demonstrates that the extrapolated \betaobs\ matches observed continuum slopes of LzLCS galaxies.

We estimated the observed, or uncorrected for internal extinction, FUV luminosity, L$_{\rm UV}$, of the galaxies as the mean flux density of the best-fit stellar model in a 100~\AA\ region centered on 1500~\AA. We then converted this into an AB absolute magnitude (\muvobs) using the redshift from the SDSS spectra.

\citet{flurya} took the ratio of the observed LyC emission to the fitted stellar continuum to estimate \fesc\ in a few ways. We predominantly use the \fesc\ derived from fitting the stellar continuum, but \fesc\ can also be derived using the observed nebular Balmer emission lines to define the intrinsic LyC emission. The different \fesc\ estimates scale with each other along a one-to-one relation, although there does exist significant scatter within the relation \citep[see figure 19 in ][]{flurya}. \citet{flurya} also derived the relative escape fraction, \fescrel, by comparing the dust-attenuated stellar continuum models to the observed LyC flux. This quantifies the LyC absorption only due to \ion{H}{i} within the galaxy. 

Finally, we inferred the production efficiency of ionizing photons as 
\begin{equation}
    \xi_{\rm ion} = \frac{Q}{L_{\rm UV}}, 
\end{equation}
where $Q$ is the intrinsic number of ionizing photons produced by the stellar population and L$_{\rm UV}$ is the intrinsic mono-chromatic UV luminosity density of the stellar population at 1500~\AA. We dereddened the best-fit stellar continuum models, divided each ionizing flux density by the respective photon energy, and integrated over the entire ionizing continuum (21--912\AA) to determine $Q$ (although the \textit{number} of hydrogen ionizing photons is dominated by photons with wavelengths near 912~\AA). We then divided $Q$ by the dereddened stellar population flux density at 1500~\AA\ from the models. The LzLCS log(\xiion~[s$^{-1}$/(erg~s$^{-1}$~Hz$^{-1}$)]) ranges from 24.94 to 25.86 with a median value of 25.47. These \xiion\ values bracket the canonical high-redshift value of 25.3 (see \autoref{reionization}), agree with those used in semi-analytical models \citep{yung20a}, and are similar to the large range of high-redshift \xiion\ values
\citep{bouwens16, harikane18, maseda, stefanon22}.

\section{THE FUV CONTINUUM SLOPE}
\label{beta}

FUV bright stellar populations are typically characterized by very blue colors because their spectral energy distributions peak in the extreme-ultraviolet \citep{claus99, raiter, eldridge17}. The blackbody nature of massive stars leads to a very steep negative power-law index at 1500~\AA\  with  \betaint\ between $-2.8$ and -2.5, after accounting for the nebular continuum, that depends on the age, metallicity, and star formation history of the stellar population \citep{claus99}. \betaint\ evolves very little over the first 10~Myr after a starburst. This is because 1550~\AA\ is still on the Rayleigh-Jeans portion of the blackbody curve for late O-stars. The non-ionizing FUV light from massive stars propagates out of the galaxy where it is absorbed and attenuated by the same gas and dust that absorbs LyC photons. This dust attenuation flattens the observed spectrum and makes \betaobs\ more positive.

\begin{figure*}
\includegraphics[width = \textwidth]{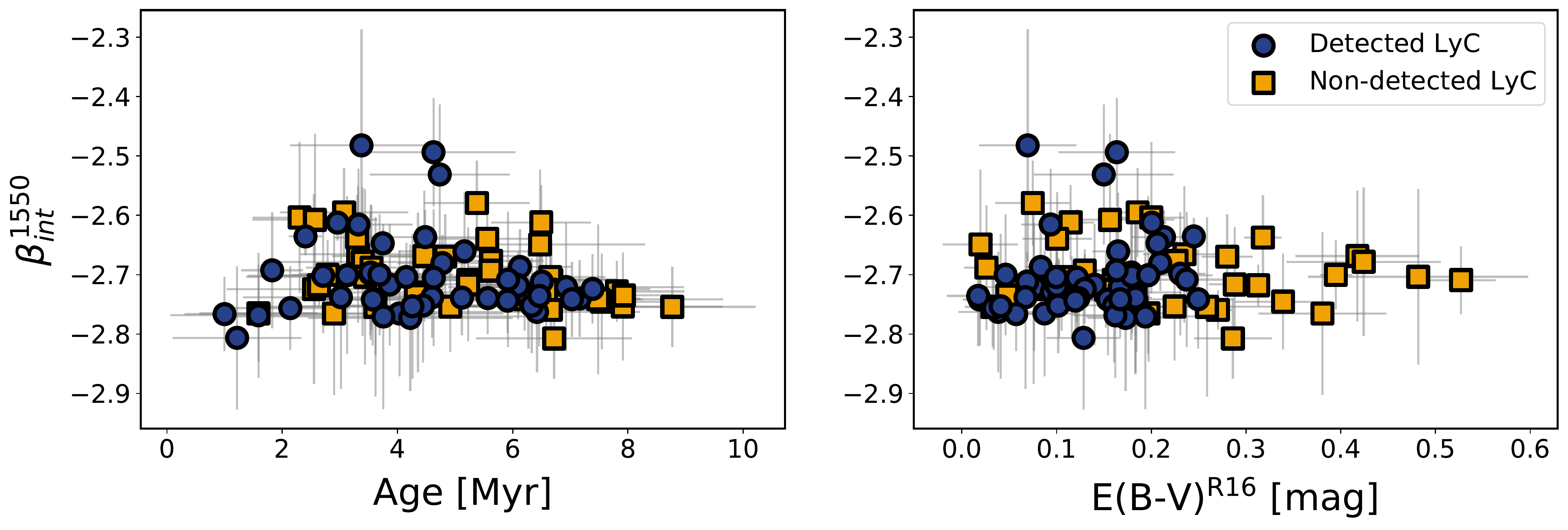}
\caption{The impact of fitted parameters on the intrinsic slope of the stellar continuum at 1550~\AA\ (\betaint) for LyC detected (blue circles) and non-detected LyC galaxies (gold squares). \textbf{Left panel: } \betaint does not scale strongly with the inferred light-weighted ages. Some scatter is introduced due to various stellar evolution that happens between 3-5 Myr. \textbf{Right Panel: } The \betaint\ is uncorrelated with the continuum color excess (\ebv$^{\rm R16}$). }
\label{fig:beta_int}
\end{figure*}
The stellar population fits allow us to explore the range of  \betaint\ within the LzLCS. The \betaint\ has a fairly narrow range of values between -2.8 and -2.5, with a median of -2.71, that does not significantly change with stellar population age (left panel of \autoref{fig:beta_int}). These values are fully consistent with the narrow range of \betaint\ expected from young stellar populations that include the nebular continuum \citep{claus99, raiter}. \autoref{fig:beta_int} shows that the \betaint\ spread does not statistically scale with age or E(B-V)$^{\rm R16}$ (Kendall's $\tau$ p-values of 0.062 and 0.37, respectively), and there is also not a statistically-consistent correlation with Z$_\ast$ (p-value of 0.065). When we refer to \ebv, we denote the attenuation law that we use to emphasize that this value depends on the assumed attenuation law (see \autoref{ebv} for how the following results can be recast to any attenuation law). This relatively narrow scatter about a single value arises from \betaint\ depending on the population age, evolutionary sequence, star formation history, and stellar metallicity (in that rough order of importance). The fits to the LzLCS continua suggest that these very young star-forming galaxies occupy  a narrow parameter range of age and star-formation history.

\begin{figure*}
\includegraphics[width = \textwidth]{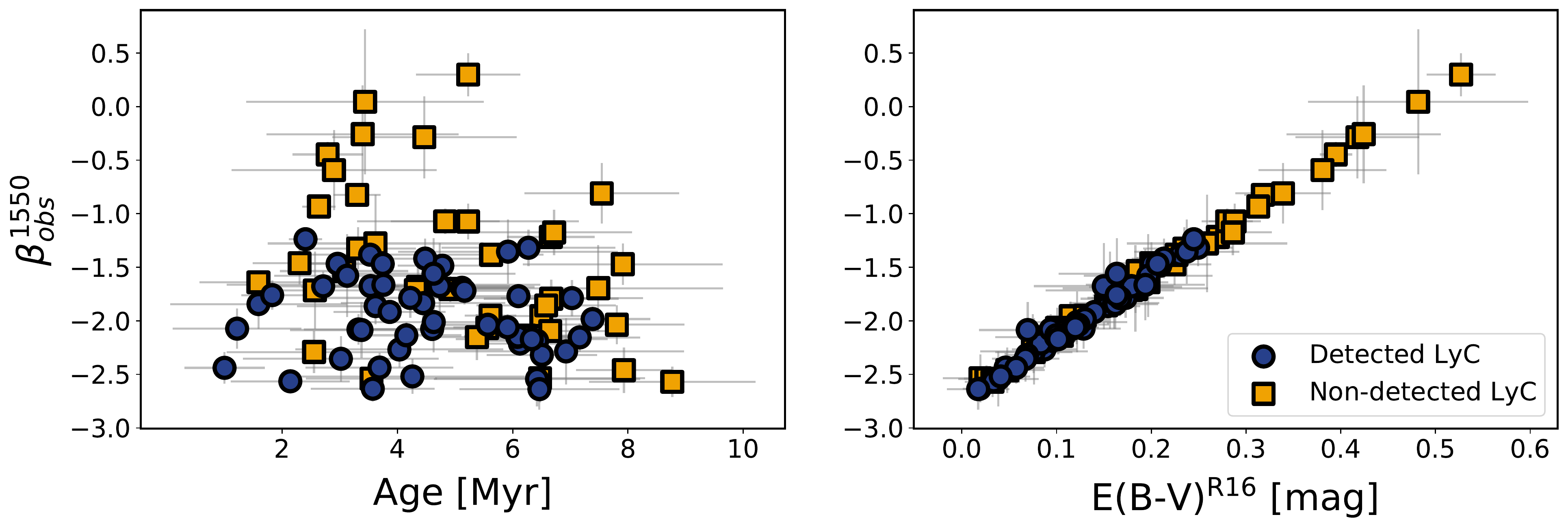}
\caption{The correlation of the observed continuum slope at 1550\AA\ (\betaobs) with fitted stellar population parameters for LyC detected (blue circles) and non-detected LyC galaxies (gold squares). \textbf{Left panel: } \betaobs\ does not strongly correlate with the light-weighted age of the stellar population. \textbf{Right panel: } The observed FUV continuum slope strongly correlates with the stellar color excess using \citet{reddy16} extinction law, \ebv$^{\rm R16}$. These two plots strongly suggest that \betaobs\ is largely set by the dust attenuation.}
\label{fig:beta_obs}
\end{figure*}

The \betaobs\ does not scale significantly with the very young light-weighted stellar population age of the LzLCS (left panel of \autoref{fig:beta_obs}), but scales strongly with the fitted \ebv$^{\rm R16}$ using a \citet{reddy16} attenuation law (right panel of \autoref{fig:beta_obs}). We use  \textsc{linmix}, a hierarchical Bayesian linear regression routine that accounts for errors on both variables \citep{linmix}, to find the relation between E(B-V)$^{\rm R16}$ and \betaobs\ is
\begin{equation}
    \text{E(B-V)}^{\rm R16} = \left(0.47 \pm 0.01\right) + \left(0.171 \pm 0.007\right) \times \beta_{\rm obs}^{1550} . \label{eq:fitted_ebv_beta}
\end{equation}
This strong correlation and small scatter suggests that there is a simple analytical relation between \ebv\ and \betaobs. If we use the assumption that the observed continuum is a power-law with wavelength (\autoref{equation:beta}) and that the observed flux is equal to the intrinsic flux times the dust attenuation factor (\autoref{eq:linear}), we find an analytic relation between \ebv\  and the continuum slope at the average of two wavelengths ($\lambda^1$ and $\lambda^2$, respectively) as 
\begin{equation}
    E(B-V) = \frac{\log_{\rm 10}{\frac{\lambda^1}{\lambda^2}}}{0.4 \Delta k} \left(\beta_{\rm int}^{<1,2>} - \beta_{\rm obs}^{<1,2>}\right) , \label{eq:beta_ebv}
\end{equation}
where $\Delta k$ is the difference in the reddening law at $\lambda^1$ and $\lambda^2$. This equation emphasizes why it is crucial to compare $\beta_{\rm obs}^{\lambda}$ values at similar wavelengths: if \ebv\ is constant, the estimated $\beta_{\rm obs}^{\lambda}$ depends on the logarithm of the ratio of the two wavelengths used to measure the spectral slope. If we use $\lambda^1 = 1300$~\AA\ and $\lambda^2 = 1800$~\AA\ (the wavelength range used to determine \betaobs), the $\Delta k$ from the \citet{reddy16} law between 1300 and 1800~\AA\ (2.08), and the median \betaint\ of -2.7, we find that \ebv\ is related to the \betaobs\ as
\begin{equation}
    E(B-V)^{\rm R16} = 0.46 + 0.17~\beta^{\rm 1550}_{\rm obs} . \label{eq:beta_ebv_numbers}
\end{equation}
This agrees with the fitted relationship between \ebv$^{\rm R16}$ and \betaobs\ (\autoref{eq:fitted_ebv_beta}). This relation will change based upon the wavelength range that \betaobs\ is determined and the attenuation law that is used (see \autoref{ebv}).

Thus, the tight correlation between \betaobs\ and \ebv\ arises due to the shape of the attenuation law and \betaint. This also explains why there is relatively little scatter within the relationship between \ebv\ and \betaobs: the extremes of the fitted \betaint\ only change the inferred \ebv\ by 0.1~mags, equivalent to the scatter of \ebv\ at fixed \betaobs\ (\autoref{fig:beta_obs}). Below, we correlate \betaobs\ with parameters like \fesc, but \autoref{eq:beta_ebv} stresses that the inferred stellar attenuation, E(B-V), does depend on how well \betaint\ can be constrained. A 0.1~mag scatter is introduced to the E(B-V) observations due to \betaint. The relatively narrow range of \betaint\ suggests that \betaobs\ is largely set by the dust reddening with a moderate impact from the \betaint. 

\begin{figure*}
\includegraphics[width = \textwidth]{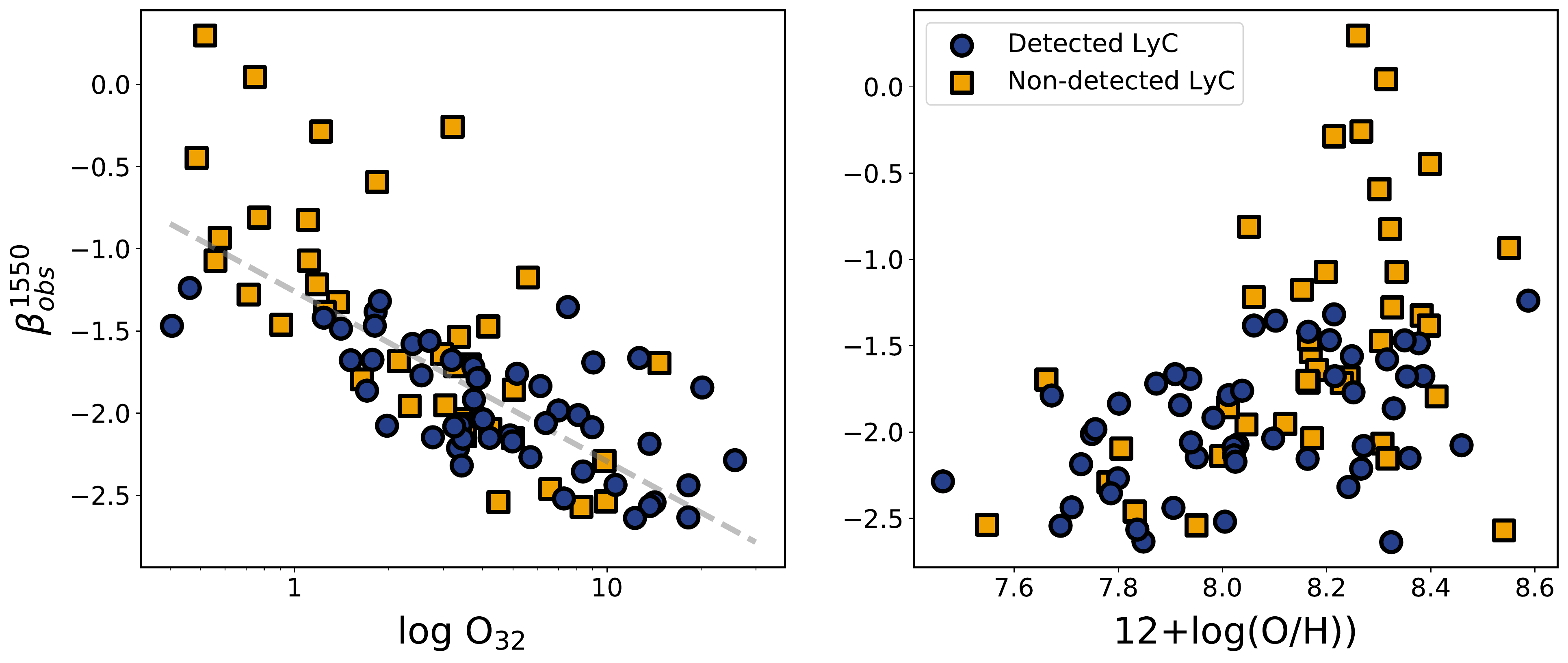}
\caption{ The observed continuum slope at 1550\AA\ (\betaobs) correlates strongly with galaxy properties derived from the optical spectra for both LyC detected (blue circles) and non-detected galaxies (gold squares). \textbf{Left panel: } \betaobs\ correlates with the flux ratio of the internal attenuation corrected [\ion{O}{iii}]~5007~\AA\ and [\ion{O}{ii}]~3727~\AA\ flux ratio  (O$_{32}$) at the 5.9$\sigma$ significance. O$_{32}$ is often used as a proxy of the ionization state of the nebular gas. Galaxies with bluer FUV slopes are more highly ionized. The gray line shows the best-fit relation between \betaobs\ and O$_{32}$ (\autoref{eq:beta_o32}). \textbf{Right Panel: } The \betaobs\ scales with the oxygen abundance (12+log(O/H)) at the 4.6$\sigma$ significance. The reddest galaxies (\betaobs~$>-0.75$) do not have direct metallicities and their 12+log(O/H) values are more uncertain. This suggests that more metal rich galaxies have redder continua and that dust attenuation increases with metallicity. }
\label{fig:o32_oh}
\end{figure*}

With this analytic understanding of \betaobs, it is useful to explore how \betaobs\ depends on other observables. The left panel of \autoref{fig:o32_oh} shows that the internal attenuation corrected O$_{32}=[\ion{O}{iii}]\text{5007~\AA}/[\ion{O}{ii}]~3727$~\AA\ flux ratio has an inverse, logarithmic, and highly significant (Kendall's $\tau$ rank coefficient of -0.55, p-value of $2\times10^{-14}$, or 7.5$\sigma$ significant) trend with \betaobs. The gray line in \autoref{fig:o32_oh} shows this best-fit relation of
\begin{equation}
    \beta^{1500}_{\rm obs} = \left(-1.3\pm0.1\right) \times \text{log}_{10} \text{O}_{32} - \left(1.0\pm0.1\right)  .
    \label{eq:beta_o32}
\end{equation}
O$_{32}$ is often quoted as an ionization parameter indicator \citep{vilchez, skillman, mcgaugh, izotov16b}, and does not have any direct observational connection to \betaobs\ (although there is a connection to the optical extinction through the Balmer dust correction). Similarly, there is a 6.7$\sigma$ (p-value of 8$\times10^{-12}$) significant correlation between \betaobs\ and the H$\beta$ equivalent width, another measure of the nebular ionization state.

\betaobs\ also scales with the gas-phase metallicity, 12+log(O/H), with a 4.6$\sigma$ significance (Kendall's $\tau = 0.34$ and p-value of $2 \times 10^{-6}$). This trend suggests that \betaobs\ becomes redder for more metal-enriched systems, and that the bluest \betaobs\ values are found in low-metallicity systems. This relation is more scattered than the O$_{32}$ and EW(H$\beta$) relations, however, all 8 of the galaxies with \betaobs\ greater than -0.75 do not have observations of the temperature sensitive [\ion{O}{iii}]~4363~\AA\ line \citep{flurya}. As such, the metallicities of the reddest galaxies may be uncertain by as much as 0.4~dex \citep{kewley06}, introducing significant scatter into this relation. Even with this significant calibration uncertainty, there is a nearly 5$\sigma$ trend between \betaobs\ and the gas-phase metallicity.

Finally, we find a 5.9$\sigma$ and 3.4$\sigma$ significant relationship between \betaobs\ and the galaxy stellar mass (\ms) and observed (uncorrected for dust attenuation) FUV absolute magnitude (\muvobs), respectively. These relations suggest that lower mass and fainter galaxies have bluer stellar continuum slopes. We fully introduce and discuss the importance of these relations in \autoref{reionization}. 

In this section we have explored relationships between both \betaint\ and \betaobs. We found that \betaint\ has a narrow range of values, but does not scale significantly with other properties. We found significant correlations between \betaobs\ and the fitted \ebv$^{\rm R16}$, the observed optical  [\ion{O}{iii}]/[\ion{O}{ii}] flux ratio, the H$\beta$ equivalent width,  \ms, \muvobs, and the gas-phase 12+log(O/H). With these correlations in mind, the next section explores the correlation between \betaobs\ and the LyC escape fraction.

\section{PREDICTING THE LYMAN CONTINUUM ESCAPE FRACTION WITH $\beta$} 
\label{fesc} 
\begin{figure}
\includegraphics[width = 0.5\textwidth]{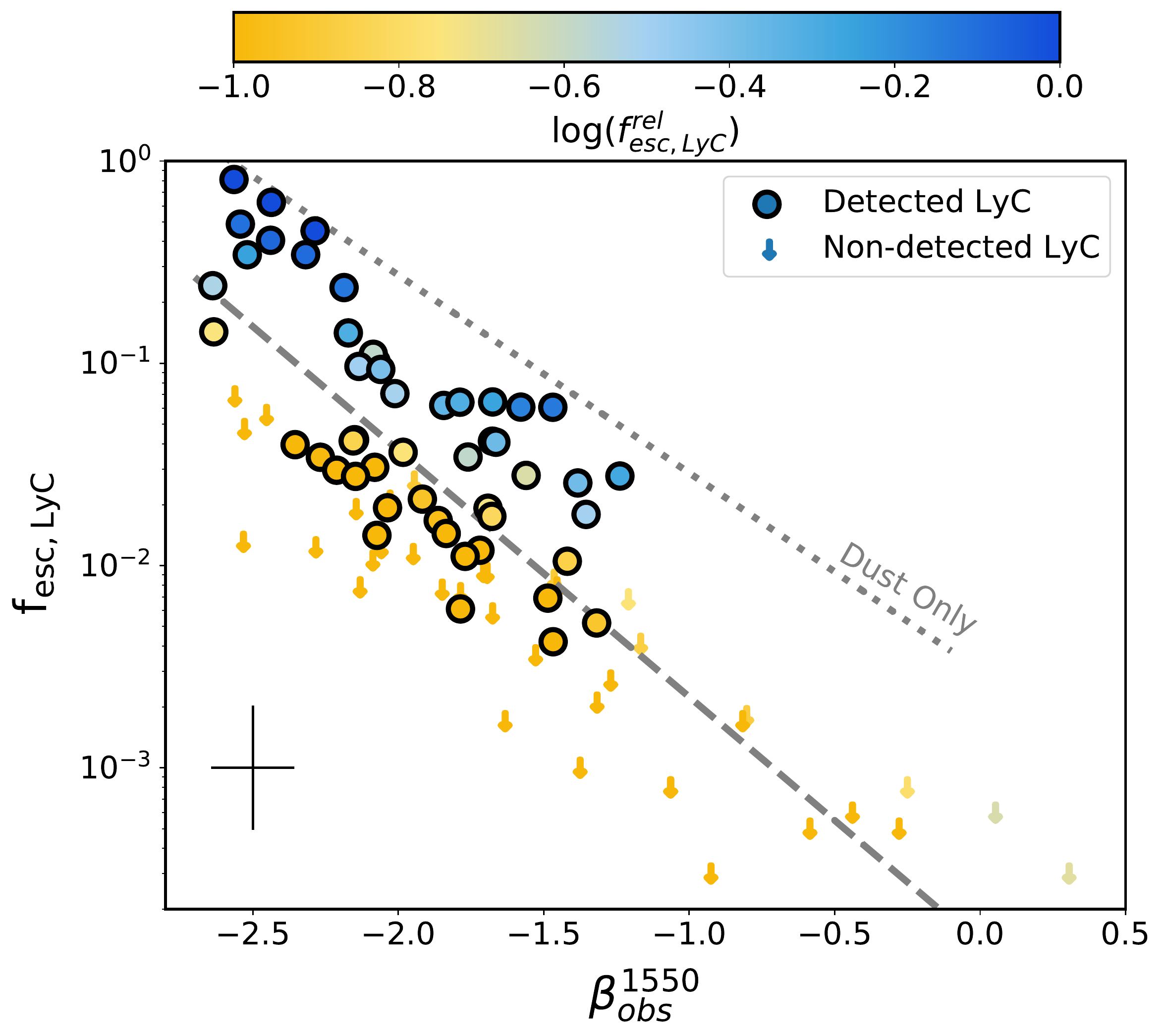}
\caption{The LyC escape fraction (\fesc) strongly scales with the observed stellar continuum slope at 1550 \AA\ (\betaobs). The LyC detections are shown as circles and the non-detections as downward pointing arrows. Representative error bars for the LyC detections are shown in the bottom left corner. Both detections and non-detections are color coded by their relative escape fractions (\fescrel), or the fraction of ionizing photons absorbed only by neutral hydrogen. At fixed \betaobs\ there is a strong vertical gradient in the color of the plotted points, suggesting that neutral hydrogen introduces significant scatter at fixed dust properties. The gray dashed line shows the best-fit relation for \fesc\ in terms of \betaobs\ (\autoref{eq:beta_fesc}) which accounts for both the \fesc\ non-detections and the errors on both \betaobs\ and \fesc. The gray dotted line shows the theoretical relation assuming that only dust attenuation removes 
ionizing photons (\autoref{eq:dust_only}). This acts as an upper envelope for the observations.}
\label{fig:beta_fesc}
\end{figure}

The previous section found that \betaobs\ scales strongly with the continuum color excess, \ebv. Similarly, \citet{Saldana} found a strong correlation between \ebv\ and \fesc. Thus, \autoref{fig:beta_obs} suggests that \betaobs\ likely correlates with \fesc\ \citep{fluryb}. \autoref{fig:beta_fesc} shows the relationship between \fesc\ and \betaobs\ with the LyC detected sample as circles and the non-detected LyC galaxies as downward pointing arrows at their corresponding 1$\sigma$ \fesc\ upper-limits.

\betaobs\ scales strongly with \fesc. The median \betaobs\ of the LyC detected sample is -2.00 and -1.63 for the non-detected sample, indicating that LyC emitting galaxies have substantially bluer continuum slopes. Using a Kendall's $\tau$ test for censored data (the $R$ package \textsc{cenken}), we find that the \betaobs\ scales with \fesc\ at the 5.7$\sigma$ significance (Kendall's $\tau = -0.42$ and a p-value of $5\times10^{-9}$). We use \textsc{linmix} \citep{linmix}, a  hierarchical Bayesian linear regression routine which accounts for the errors on both \fesc\ and \betaobs\ as well as for the upper limits on \fesc,  to find the analytic relation between \fesc\ and \betaobs\ (the gray dashed line on \autoref{fig:beta_fesc})  of
\begin{equation}
    f_{\rm esc, LyC} = \left(1.3 \pm 0.6 \right) \times 10^{-4} \times 10^{\left(-1.22\pm0.1\right) \beta^{1550}_{\rm obs}} . \label{eq:beta_fesc}
\end{equation}
Many theoretical models for reionization suggest that star-forming galaxies must emit more than 5, 10, or 20\% of their ionizing photons to reionize the IGM \citep{ouchi, robertson13, robertson15,rosdahl18, finkelstein15, finkelstein19, naidu19}. \autoref{eq:beta_fesc} suggests that these \fesc\ occur on average in the LzLCS for \betaobs\ values less than $-2.11$, $-2.35$, and $-2.60$, respectively. We also find a 2.9 and 3.4$\sigma$ significant correlation (p-values of 0.002 and 0.0003, respectively) between \betaobs\ and \fescrel\ and the escape fraction calculated using the H$\beta$ equivalent width \citep{flurya}. Thus, \betaobs\ correlates with complementary \fesc\ measures.  

 \autoref{eq:beta_fesc} describes the population average relation between \betaobs\ and \fesc. This relation contains real and significant scatter that must be addressed when applying it to higher redshift observations. With the appreciable scatter in this relation, \autoref{eq:beta_fesc} more robustly estimates population averaged escape fractions rather than individual galaxy-by-galaxy escape fractions. How much of the scatter is physical? The LzLCS \fesc\ detections have a median signal-to-noise of 3 \citep{flurya}. Thus, there are significant observational uncertainties in the \fesc\ measurements (see the representative error bars in the lower left of  \autoref{fig:beta_fesc}). \textsc{linmix} estimates that about 0.15~dex of the scatter in \autoref{fig:beta_fesc} is intrinsic to the relationship.  A possible source of intrinsic scatter is the neutral gas column density, the neutral gas covering fraction, or the geometry. These have been observed to scale strongly with \fesc\ in the LzLCS \citep{Saldana}. The points in \autoref{fig:beta_fesc} are color-coded by \fescrel\ or the escape fraction in the absence of dust. While \fescrel\ only moderately scales with \betaobs\ (3.4$\sigma$), at fixed \betaobs\ there is a strong vertical \fescrel\ gradient, which may suggest that galaxy-to-galaxy variations in neutral gas sets the scatter in \autoref{fig:beta_fesc}. This strong secondary scaling with \fescrel\ echoes the findings in \autoref{beta} where \betaobs\ strongly scales with O$_{32}$ and H$\beta$ EW, both which trace the ionization state.  In a future paper (Jaskot et al. in preparation), we will explore multivariate correlations with \fesc\ and \betaobs, O$_{32}$, H$\beta$ equivalent width, metallicity, or star formation rate surface density.  

\begin{figure}
\includegraphics[width = 0.5\textwidth]{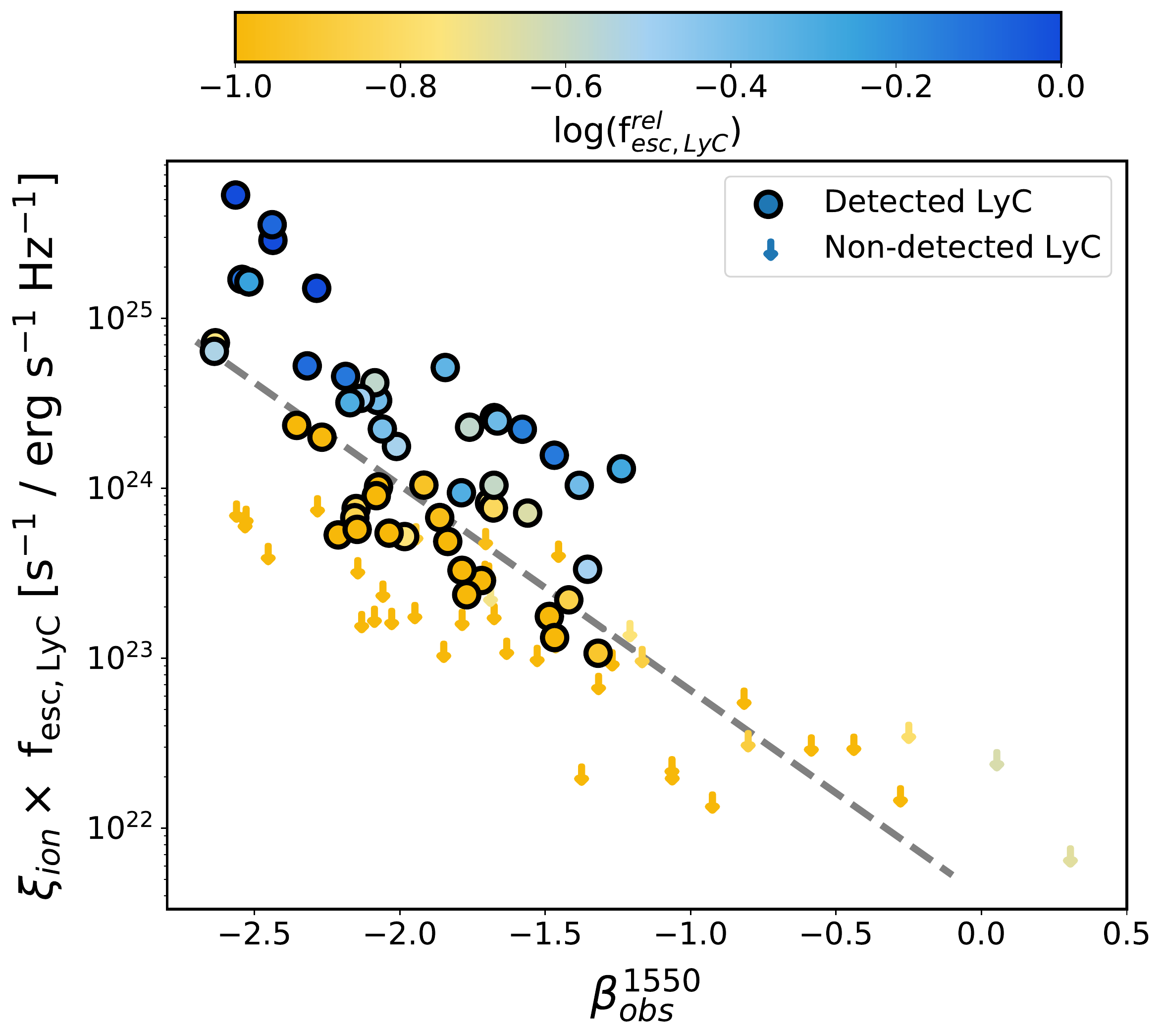}
\caption{The LyC escape fraction (\fesc) times the ionizing photon production efficiency ($\xi_{\rm ion}$ = Q/L$_{\rm UV}$) strongly correlates with the observed continuum slope at 1550 \AA\ (\betaobs). The gray dashed line shows the best-fit relation for \fesc~$\times$~$\xi_{\rm ion}$ in terms of \betaobs\ (\autoref{eq:beta_fesc_xi}) that accounts for both the \fesc\ upper-limits and errors.}
\label{fig:beta_fesc_xi}
\end{figure}

Finally, \autoref{fig:beta_fesc_xi} shows the scaling relation between \betaobs\ and the emitted ionizing efficiency ($\xi_{\rm ion} \times f_{\rm esc}$). Using \textsc{linmix} we determine this relation to be 
\begin{equation}
    \xi_{\rm ion} \times f_{\rm esc} = (4 \pm 2) \times 10^{21} \text{[s$^{-1}$ / erg s$^{-1}$ Hz$^{-1}$]} \times 10^{\left(-1.2\pm0.1\right) \text{\betaobs}} .
    \label{eq:beta_fesc_xi}
\end{equation}
This has a similar \betaobs\ dependence as the \fesc\ relation given in \autoref{eq:beta_fesc} with a power law exponent of -1.2. The scatter on the emitted ionizing efficiency relation is similar to the scatter in \autoref{eq:beta_fesc}, consistent with the insignificant relationship found between \betaobs\ and \xiion\ (p-value of 0.155). Thus, the ionizing emissivity scales the relation between \betaobs\ and \fesc\ to include the production efficiency of ionizing photons. In \autoref{reionization} we use \autoref{eq:beta_fesc_xi} to estimate the ionizing emissivity of galaxies during the epoch of reionization. 

\section{IMPLICATIONS FOR THE ESCAPE OF IONIZING PHOTONS}
\label{escape}

The FUV spectral slope strongly correlates with \fesc. \autoref{fig:beta_int} and \autoref{fig:beta_obs} demonstrate that the major determinant of the FUV spectral slope is the dust attenuation derived from the observed continuum shape. Dust connects the gas-phase metallicity through the dust-to-gas ratio, and to the ionization state of the gas because metals are the predominate nebular coolant. The metallicity of the gas strongly depends on the stellar mass of the galaxies through the mass-metallicity relation \citep{tremonti04, berg} and the mass-luminosity relation. Therefore, \betaobs, \ms, O$_{32}$, \muvobs, and \ebv\ are all linked together through the gas-phase metallicity. It is not entirely surprising then that \autoref{beta} found that all of these parameters correlate. This connection between LyC and dust echoes relations between \lya\ and dust \citep{hayes11}. Here we explore the physical connection between \betaobs\ and \fesc\ and how the observed \betaobs-\fesc\ relation informs on how ionizing photons escape galaxies. 

To explore the connection between \betaobs\ and \fesc, we revisit the absolute escape fraction definition \citep[see][]{Saldana}. There are two main sinks of ionizing photons: dust (with an optical depth of $\tau_{\rm dust}$) and \ion{H}{i} \citep{chisholm18}. Thus, \fesc\ is determined as the product of the attenuation of dust and \fescrel\ as
\begin{equation}
    f_{\rm esc, LyC} = 10^{-0.4\tau_{\rm dust}} \times f_{\rm esc,  LyC}^{\rm rel}\label{eq:radiative}~,
\end{equation}
where 
\begin{equation}
    \tau_{\rm dust}(912) = k(912)~{\rm E(B-V)} . 
\end{equation}
A possible explanation for the significant correlation between \betaobs\ and \fesc\ is that dust is the only sink of ionizing photons. This can be tested by assuming that the \ion{H}{i} transmits all of the ionizing photons (\fescrel~=~1). \autoref{eq:radiative} then only depends on the dust as
\begin{equation}
    \log(f_{\rm esc, LyC}^{\rm dust}) \approx -0.4\tau_{\rm dust}(912) = -0.4E(B-V)k(912)  .
\end{equation}
Using \autoref{eq:beta_ebv}, \autoref{eq:beta_ebv_numbers}, and the fact that k(912)=12.87 for the \citet{reddy16} extinction law, this equation can be redefined in terms of \betaobs\ as
\begin{equation}
    \log(f_{\rm esc, LyC}^{\rm dust}) \approx -k(912) \frac{\log_{\rm 10}{\frac{\lambda^1}{\lambda^2}}}{\Delta k} \left(\beta_{\rm int} - \beta_{\rm obs}\right) = -2.4 -0.9\beta^{1550}_{\rm obs} . \label{eq:dust_only}
\end{equation}
We overplot this relation as the dotted line in \autoref{fig:beta_fesc} and see that this relation provides a strong upper envelope for  \fesc. There are no observed points above this \lq{}\lq{}dust only\rq{}\rq{} line. All the points lie below the \lq{}\lq{}dust only\rq{}\rq{} line because neutral gas also decreases \fesc. As emphasized in \autoref{eq:radiative}, \fesc\ is the product of the dust and \fescrel.  This naturally explains why there is a strong gradient of \fescrel\ in \autoref{fig:beta_fesc} at fixed \betaobs: the absolute escape fraction is a product of both dust and \ion{H}{i} opacity.  \citet{Saldana} used the Lyman Series absorption lines to find that the points close to the \lq{}\lq{}dust only\rq{}\rq{} line have very low neutral gas covering fractions and \ion{H}{i} absorbs little LyC, explaining why \fescrel~$\sim 100$\% for these points. 

\section{IMPLICATIONS FOR COSMIC REIONIZATION}
\label{reionization}
The goal of the LzLCS is to provide local examples of LyC escape that are  directly transferable to high-redshift observations to determine their contribution to reionization. To reionize the early IGM, star-forming galaxies must have an ionizing emissivity, $\dot{n}_{\rm ion}$, much larger than the cosmic baryon density at high-redshift. \autoref{eq:nion} infers $\dot{n}_{\rm ion}$ if the integrated FUV luminosity function ($\rho_{\rm UV}$), the ionizing efficiency ($\xi_{\rm ion})$, and \fesc\ are observed \citep{ouchi, robertson13, robertson15, finkelstein15, finkelstein19, mason19, naidu19}. Above, we have provided an empirical route to estimate \fesc\ and \xiion$\times$\fesc\ using the FUV continuum slope, a common observable of high-redshift galaxies  \citep{finkelstein12, dunlop13, bouwens14, wilkins16, roberts21, Bhatawdekar21, sandro21}. If we assume that these LzLCS relations apply to high-redshift galaxies, do early star-forming galaxies emit sufficient ionizing photons to reionize the early IGM?

\begin{figure}
\includegraphics[width = 0.5\textwidth]{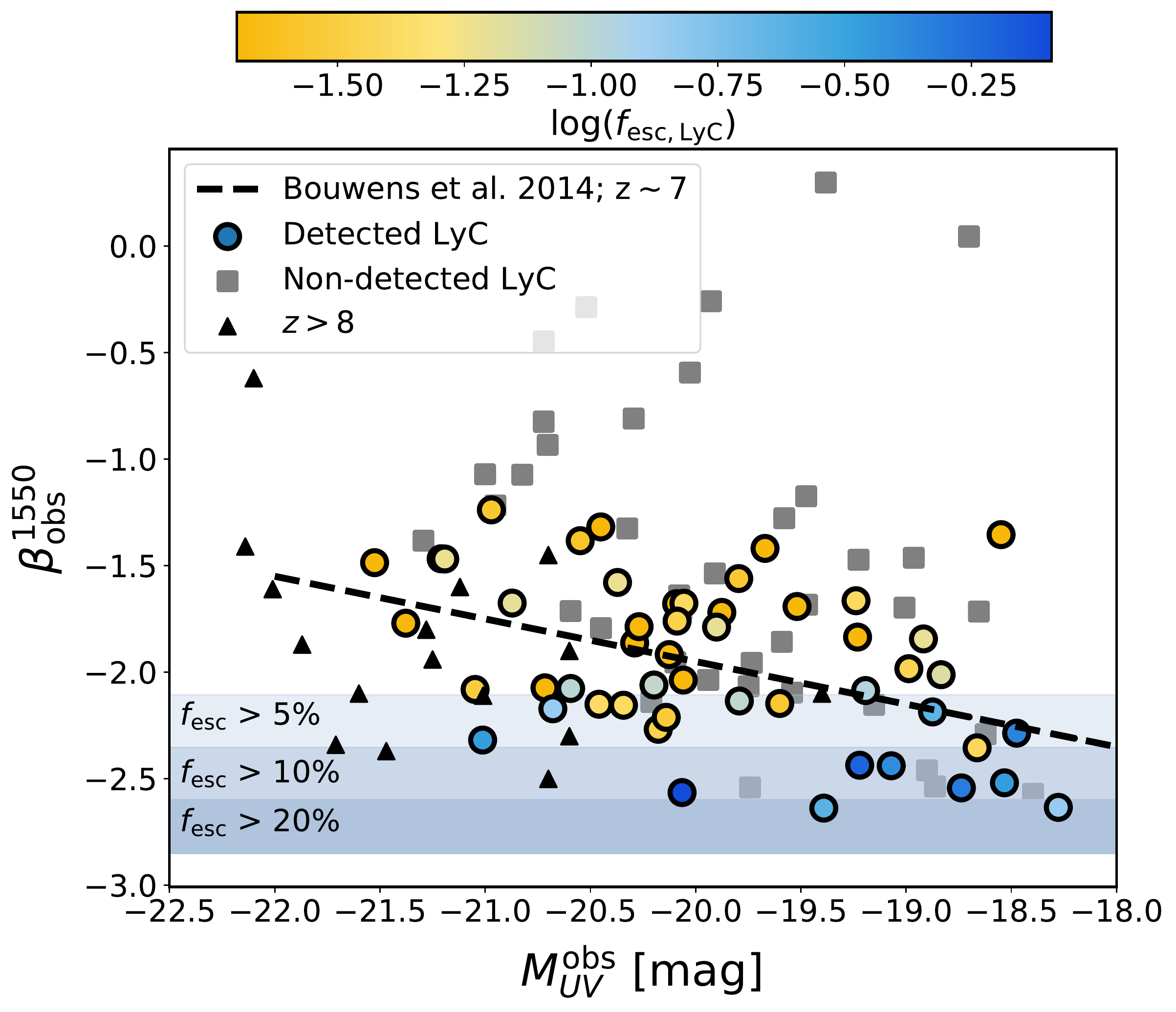}
\caption{The observed absolute FUV magnitude (\muvobs) of the LyC detections (colored circles) and non-detections (gray squares) for the LzLCS sample scales with the observed stellar continuum slope at 1550~\AA\ (\betaobs) at the 3.4$\sigma$ significance. Brighter LzLCS galaxies are redder and possibly emit a smaller fraction of their ionizing photons. The \betaobs-\muvobs\ observed at $z\sim 7$ from \citet{bouwens14} is included as a dashed line. This $z \sim 7$ relation matches the general trend of the LyC detections in the LzLCS. Black triangles show individual $z > 8$ galaxies from \citet{wilkins16}, \citet{Bhatawdekar21}, and \citet{sandro21}. The LyC detections are colored by the observed LyC escape fraction (log(\fesc)) to emphasize galaxies with \fesc~$>$5\% as blue points. We use \autoref{eq:beta_fesc} and \betaobs\ to estimate population averages  of \fesc\ equal to 5, 10, and 20\% . In general, fainter galaxies have larger \fesc.}
\label{fig:beta_muv}
\end{figure}

\autoref{fig:beta_muv} compares \betaobs\ to the LzLCS observed (not corrected for dust) \muvobs. We color-code the LzLCS points as gray squares for non-detections and detected galaxies with the observed \fesc. The colorbar is shifted such that blue points represent so called \lq{}\lq{}cosmologically relevant\rq{}\rq{} escape fractions ($>$5\%). Using the relationship between \betaobs\ and \fesc\ (\autoref{eq:beta_fesc}) we shade regions of this figure that correspond to LzLCS population averages of 5, 10, and 20\% escape fractions. This shading emphasizes that the LzLCS galaxies with \fesc\ greater than 5\% are predominately found at fainter \muvobs\ \citep{fluryb}.

\begin{figure}
\includegraphics[width = 0.5\textwidth]{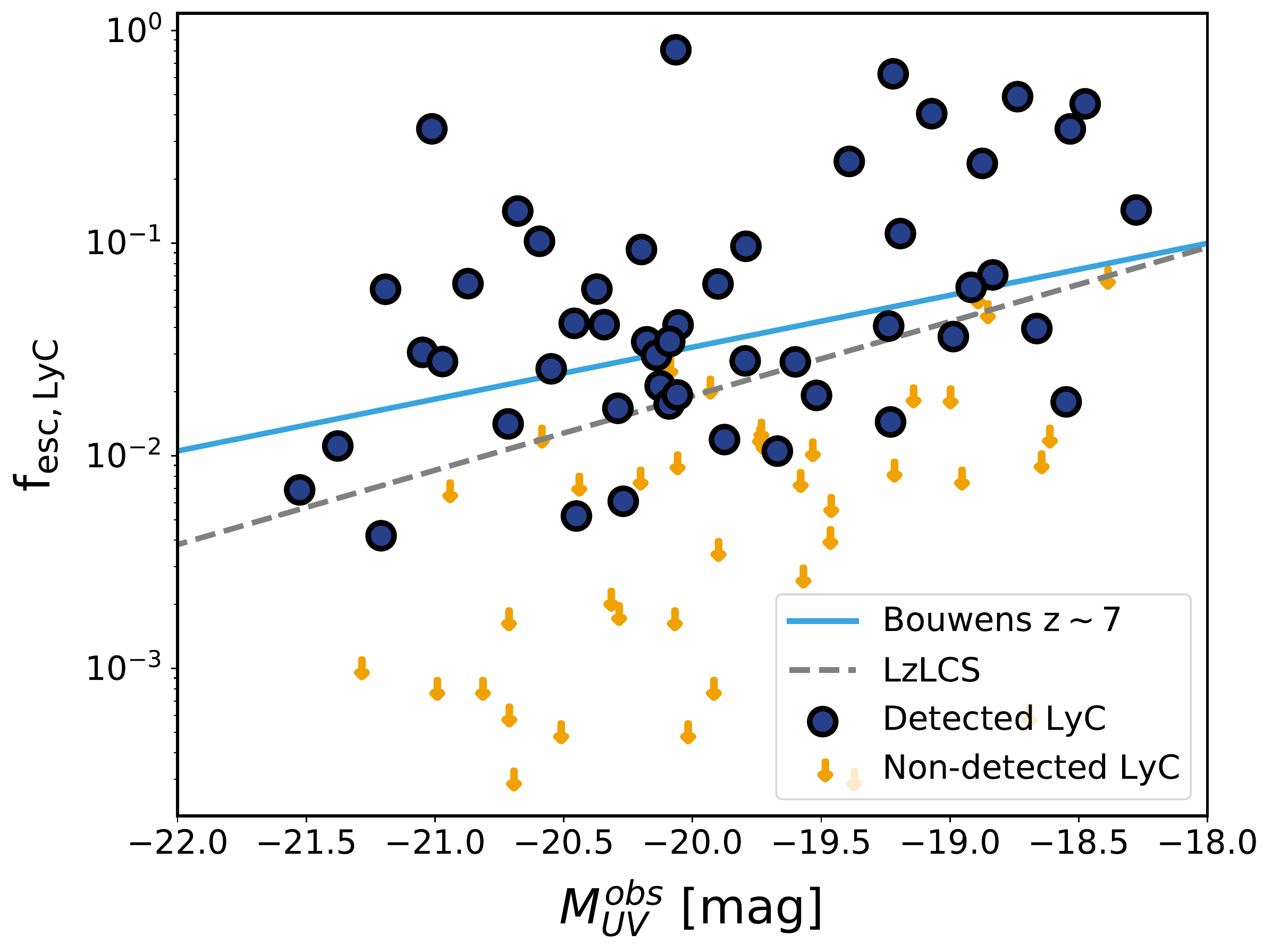}
\caption{The observed absolute FUV magnitude (\muvobs) versus the observed LyC escape fraction (\fesc) for galaxies with detected (blue circles) and non-detected (downward pointing arrows) LyC emission. The inferred relationship at $z\sim7$ is shown as a light blue line using \autoref{eq:bouwens_fesc}  and the observed relation between \betaobs\ and \muvobs\  from \citet{bouwens14}. The best-fit relation from the LzLCS, accounting for the upper limits on \fesc, is shown as the dashed gray line. The LzLCS and $z\sim7$ relations have statistically similar slopes ($0.34\pm0.10$ and $0.25\pm0.09$ for the LzLCS and $z\sim7$ relation, respectively) and normalizations ($5.27\pm2.1$ and $5.14\pm0.3$ respectively).  }
\label{fig:fesc_MUV_obs}
\end{figure}

Many previous studies have explored the median relationship  between  \betaobs\ and \muvobs\ at high-redshift \citep{finkelstein12, dunlop13, bouwens14}. \citet{bouwens14} fit the \betaobs-\muvobs\ relation at different redshifts as 
\begin{equation}
    \beta^{1550}_{\rm obs} = b + a (M_{\rm UV}^{\rm obs} +19.5) . \label{eq:beta_bouwens} 
\end{equation}
Table 3 in \citet{bouwens14} gives the $a$ and $b$ values for six different redshifts from 2.8 -- 8.  We note that the $z = 8$  \betaobs-\muvobs\ relation from \citet{bouwens14} has a fixed slope ($a$ value) due to the uncertain $Y$-band data. The dashed line in \autoref{fig:beta_muv} shows the $z=7$ relation over the LzLCS observations. The high-redshift relation matches the LzLCS across most \muvobs. This suggests that the LzLCS and $z\sim7$ galaxies have a similar relation between their FUV brightnesses and continuum slopes. Combining the observed $z\sim7$ \betaobs-\muvobs\ relation (\autoref{eq:beta_bouwens}) and the LzLCS \betaobs-\fesc\ relation (\autoref{eq:beta_fesc}), we predict the population-averaged \fesc\ at various redshifts as 
\begin{equation}
    f_{\rm esc,LyC} = 1.3\times10^{-4} \times 10^{-1.22 \left(b+a\left[M_{\rm UV}^{\rm obs} + 19.5\right]\right)} . 
    \label{eq:bouwens_fesc}
\end{equation}
In \autoref{fig:fesc_MUV_obs}, we overplot this predicted $z\sim7$ relationship  using the $a$ and $b$ values from \citet{bouwens14} as a light blue line. The dashed gray line illustrates the best-fit LzLCS relation using \textsc{linmix} \citep{linmix} to account for the \fesc\ upper limits and the errors on both \fesc\ and \muvobs. The LzLCS and $z\sim7$ relations have relatively large error bars, but their slopes and normalizations are statistically similar (a slope in log-space of $0.34\pm0.10$ and $0.25\pm0.09$ for the LzLCS and $z\sim7$ relations, respectively). Thus, the $z\sim7$ relation roughly reproduces the observed LzLCS trend. The observed log-space trend is very shallow \citep{fluryb} and suggests only a factor of 10 \fesc\ change in over 4 orders of magnitude of \muvobs. The large LzLCS \muvobs\ dynamic range likely explains why smaller previous samples did not find a statistically significant trend between \muvobs\ and \fesc\ \citep{izotov21}.

Using the \citet{bouwens14} observations, \autoref{eq:bouwens_fesc} predicts that $z\sim7$ galaxies emit $>5$\% ($>20$\%) of their ionizing photons if they are fainter than \muvobs~$>$~-19.2~mag ($-16.7$~mag; see the shading in \autoref{fig:beta_muv}). Other works have made similar \betaobs\ observations at high-redshift. \citet{dunlop13} find \betaobs~=~$-2.08$ at $z\sim7$ for \muvobs~=~$-18.5$~mag galaxies and $-1.81$ for \muvobs~=~$-19.5$~mag galaxies. This leads to a population average \fesc~$\simeq$~5\% for faint galaxies, while brighter galaxies have an \fesc\ of 2\%. The triangles in \autoref{fig:beta_muv} show individual $z>8$ galaxies \citep{wilkins16, Bhatawdekar21, sandro21}. While sparsely sampled at the moment, $z > 8$ galaxies fainter than $-20.5$~mag have a median \betaobs~=~$-2.1$ (suggesting an \fesc\ near 5\%) and galaxies brighter than $-20.5$ have \betaobs~=~$-1.9$. Since \betaobs\ largely tracks the dust contribution to \fesc\ (\autoref{escape}), we can use \autoref{eq:dust_only} to put an upper limit of \fesc~$<30$\% for the individual \muvobs~$>$~$-20.5$ $z\sim8$ galaxies. This is the \lq{}\lq{}dust free\rq{}\rq{} scenario and modest \fescrel\ can reduce \fesc\ to population averages of 5\%. Finally, we note that the LzLCS \betaobs-\muvobs\ relations are entirely consistent with semi-analytical models of \betaobs\ at high-redshift, including the large scatter to higher \betaobs\ values \citep{yung19}. 

\begin{figure}
\includegraphics[width = 0.5\textwidth]{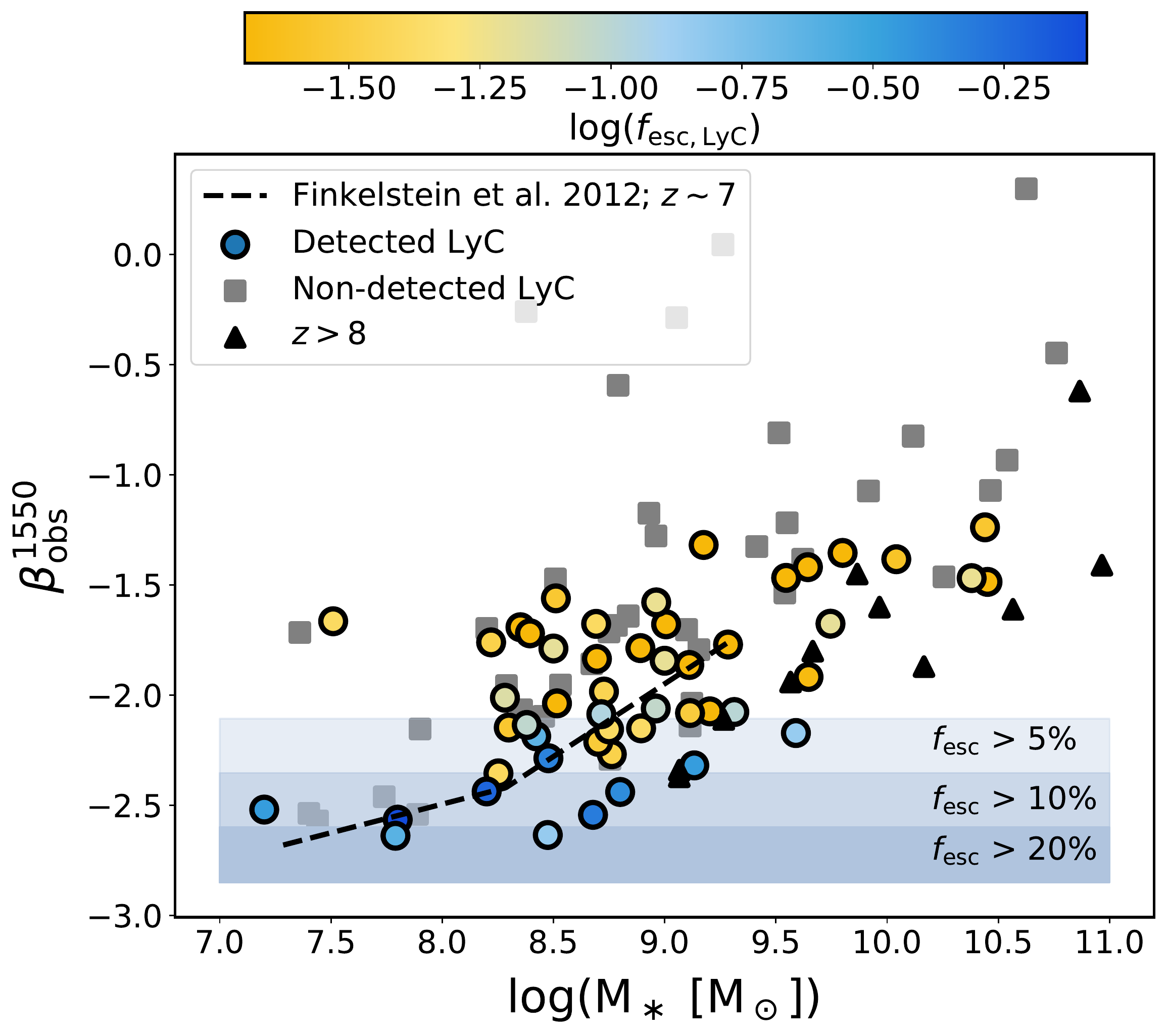}
\caption{ LzLCS galaxies that are blue and low stellar mass (\ms) tend to have larger LyC escape fractions. The \betaobs\ of the LzLCs detections (colored circles) and non-detections (gray squares) scale with the \ms\ at the 5.9$\sigma$ significance. We include the $z \sim 7$ relation from \citet{finkelstein12} as the dashed line and individual $z > 8$ galaxies from \citet{wilkins16}, \citet{Bhatawdekar21}, and \citet{sandro21} as black triangles. The LzLCS LyC detections are colored by the observed LyC escape fraction (log(\fesc)), such that galaxies with \fesc~$>$~5\% are blue. We use \autoref{eq:beta_fesc} to shade regions of the plot that correspond to 5, 10, and 20\% \fesc. The overall trend suggested by the LzLCS for the high-redshift galaxies is that lower \ms\ galaxies have higher \fesc. }
\label{fig:beta_ms}
\end{figure}

The stellar mass, \ms, represents the integrated star formation of a given galaxy.  \autoref{fig:beta_ms} shows the 5.9$\sigma$ significant relationship between \ms\ and \betaobs. More massive galaxies have redder continuum slopes than their lower mass counterparts, likely as their gas-phase metallicity \citep{tremonti04, mannucci, berg} and dust content increases \citep{remy13, popping17, popping22}. The relationship between \betaobs\ and \ms\ for a sample of $z \sim 7$ galaxies from \citet{finkelstein12} is overplotted as a black dashed line in \autoref{fig:beta_ms}. We choose the $z \sim 7$ range from \citet{finkelstein12}  because $z\sim7$ is squarely within the epoch of reionization and it is the same redshift used for \muvobs\ from \citet{bouwens14}. Similar to \autoref{fig:beta_muv}, the correspondence between the high redshift samples and the LzLCS is striking: the \citet{finkelstein12} relation tracks through the center of the LzLCS points (both detections and non-detections). This suggests that the LzLCS \betaobs\ varies with host galaxy properties in similar ways as high-redshift galaxies. We use \autoref{eq:beta_fesc} to shade regions in \autoref{fig:beta_ms} of expected \fesc\ values.  On average, $z \sim 7$ galaxies with log\ms$< 8.7$ (7.6) emit more than 5\% (20\%) of their ionizing photons. \autoref{fig:beta_ms} also includes  individual $z > 8$ galaxies from the literature \citep{wilkins16, Bhatawdekar21, sandro21}. Of all the individual $z > 8$ galaxies, only GOODSN-35589 and EGS-68560, both at log\ms~=~9.1, are blue enough to emit more than 5\% of their ionizing photons. These galaxies are weak LyC emitters with a predicted \fesc~$\sim$~6\%. The other $z >8$ sources are too red to emit appreciable LyC. This is likely because most of the individual literature sources are biased towards bright and massive galaxies. If the LzLCS relationship between \betaobs\ and \fesc\ holds for galaxies within the epoch of reionization, faint and low stellar mass galaxies likely dominate the ionizing photon budget. 

\begin{figure}
\includegraphics[width = 0.5\textwidth]{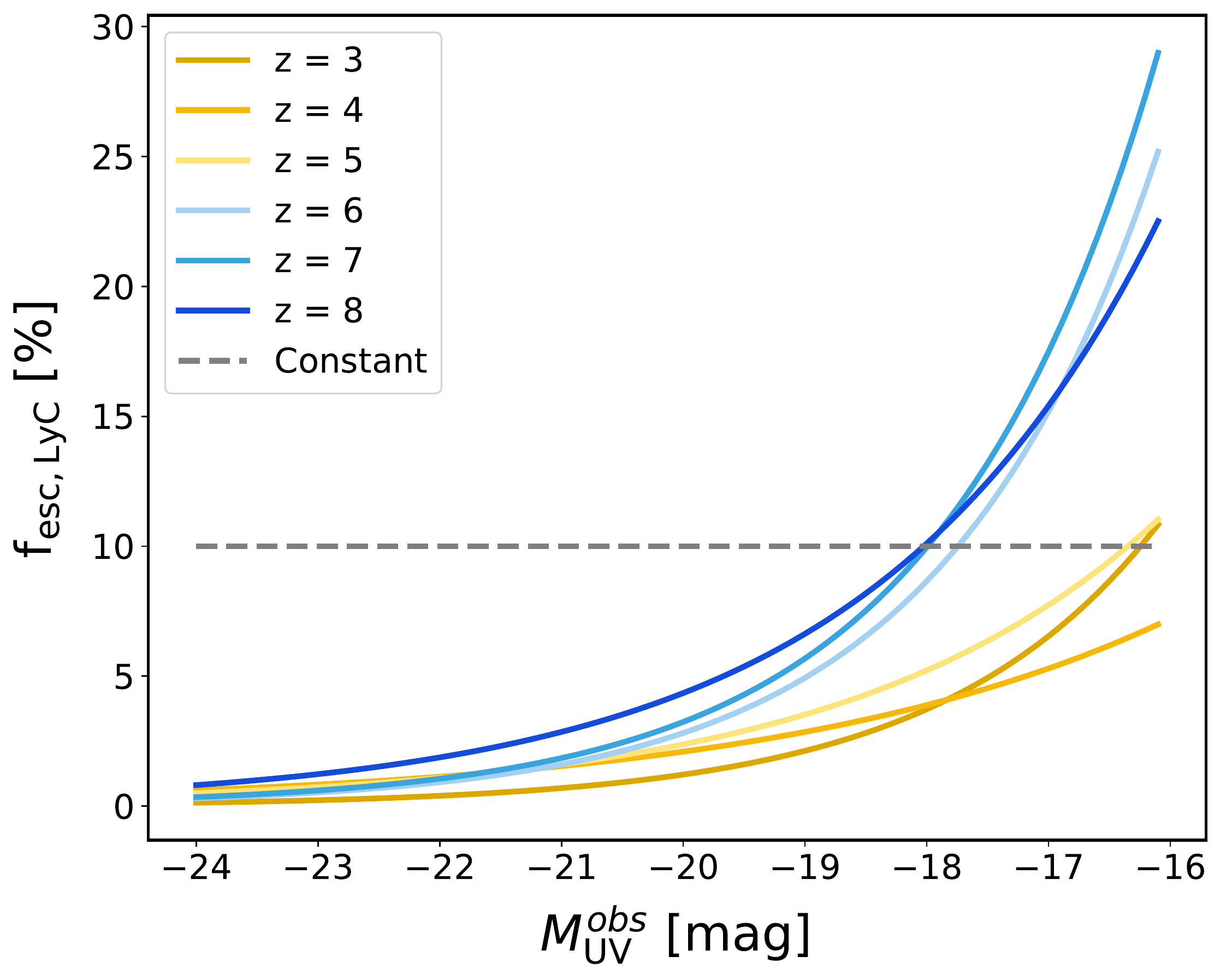}
\caption{The inferred evolution of \fesc\ at high-redshift, assuming the relation between \betaobs\ and \fesc\ (\autoref{fig:beta_fesc}) applies for the \muvobs-\betaobs\ relations from \citet{bouwens14} (\autoref{eq:beta_bouwens}). We estimate \fesc\ at 6 different redshifts and provide an estimate of the evolution of \fesc\ with \muvobs. The \fesc\ varies shallowly with \muvobs\ for galaxies brighter than $-20$~mag. In general, these observationally-motivated relations suggest that fainter galaxies at higher redshifts emit a larger fraction of their ionizing photons.} 
\label{fig:fesc_muv}
\end{figure}

The LzLCS relations and the high-redshift \betaobs\ observations allow for some of the first indirect estimates of \fesc\ during the epoch of reionization. We convert the \citet{bouwens14} \betaobs\ relations in \autoref{eq:bouwens_fesc} into population averages of \fesc\ using the LzLCS relation. The predicted \fesc\ curves for \muvobs\ down to -16~mag, often the current detection limits, are plotted in \autoref{fig:fesc_muv} for 6 different redshifts. These observationally motivated relations suggest that only galaxies fainter than -16~mag have \fesc\ $>$10\% at $z < 5$. This is broadly consistent with current samples of LyC emitters at $z\sim3$: galaxies brighter than L$^{\ast}_{z=3}$ (\muvobs~=~-21~mag) have non-detected LyC, while \fesc\ increases with decreasing luminosity and is \fesc~$\sim$~12\% for galaxies fainter than L$^\ast_{z=3}$  \citep{pahl21}.  During the epoch of reionization (blue curves), all galaxies brighter than L$^\ast_{z=3}$ (\muvobs~$\lesssim -21$~mag) have population-averaged escape fractions less than 5\%. \fesc\ increases for fainter galaxies as their  \betaobs\ becomes bluer. During the epoch of reionization, the population averaged \fesc\ exceeds 10\% for \muvobs\ values fainter than -18~mag. These \fesc\ trends are in slight tension with recent \fesc\ estimates using the \lya\ luminosity function \citep{matthee22}. This work found the $z\sim7$ \fesc\ distribution to peak near \fesc~=~11\% for galaxies with \muvobs~=~-19.5 (a value two times higher than predicted in \autoref{fig:fesc_muv}) and decreases for both brighter and fainter galaxies. The difference could be due to a \betaobs\ tracing the evolving metal content of galaxies while \citet{matthee22} determine \fesc\ from the \lya\ properties.

Can metallicity evolution explain the dramatic increase of \fesc\ from $z = 5$ to $z = 6$ observed in \autoref{fig:fesc_muv} as reionization concludes? The increase in \fesc\ at $z > 6$ for faint galaxies in \autoref{fig:fesc_muv} is due to the $b$ value in \autoref{eq:beta_bouwens} from \citet{bouwens14} becoming more negative. The $b$ measures the \betaobs\ zero-point at fixed \muvobs. As $b$ becomes more negative, from $-1.9\pm0.02$ at $z=5$ to $-2.13\pm0.44$ at $z = 8$, galaxies become bluer at fixed \muvobs. What causes this blue shift?  \autoref{fig:o32_oh} and \autoref{fig:beta_ms} show that changes to gas-phase ionization, metallicity, and \ms\ heavily impacts \betaobs. Galaxies that vigorously form stars (and simultaneously produce ionizing photons), also rapidly synthesize metals. As galaxies grow in \ms\ they retain more of the these metals \citep{tremonti04}, allowing for them to create more dust \citep{remy13, popping17, popping22}. Thus, the rapid buildup of stellar mass and metallicity during the epoch of reionization could lead to redder \betaobs\ and push \fesc\ to lower values at later times. 

\begin{figure}
\includegraphics[width = 0.5\textwidth]{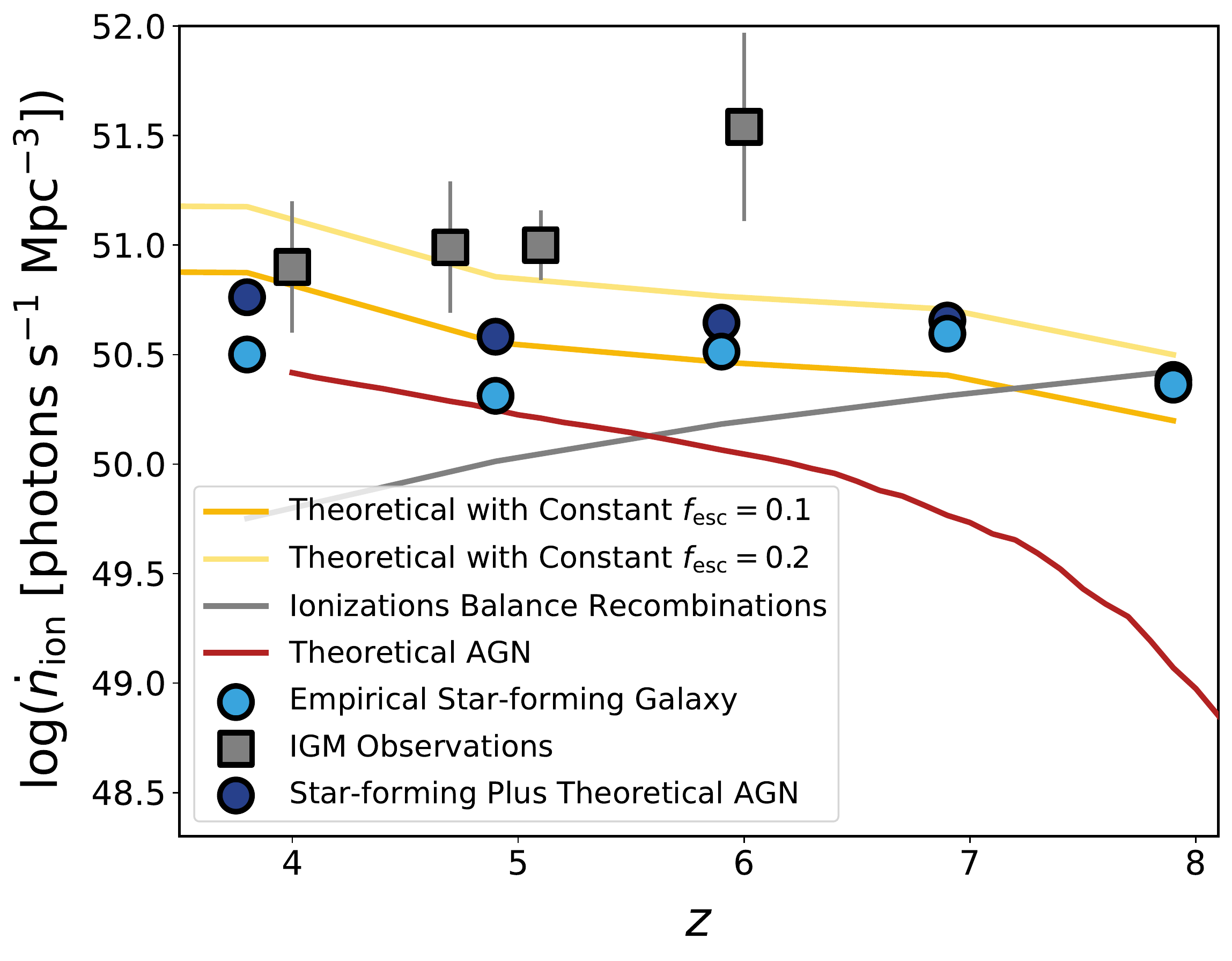}
\caption{An observationally-motivated estimate of the ionizing emissivity (\nion) during the epoch of reionization. The light blue circles are the values predicted using the LzLCS relation between \betaobs\ and \xiion$\times$\fesc\ (\autoref{eq:beta_fesc_xi}) and the observed \muvobs-\betaobs\ relation (\autoref{eq:beta_bouwens}) at redshifts 4, 5, 6, 7, and 8 from \citet{bouwens14}. The dark and light gold curves are estimates assuming a constant \fesc\ of 10 and 20\%, respectively. The gray squares are observations of the ionizing emissivity \citep{becker13, becker21}. The gray line is the ionizing emissivity required for ionization to balance recombination at a given redshift \citep{madau99, ouchi}, assuming a redshift-dependent clumping factor \citep{shull12}. Values above the gray line will increase the hydrogen ionization fraction in the IGM. The red line is a theoretical estimate of the AGN emissivity from \citet{finkelstein19}. The dark blue circles are the sum of the AGN and the empirical star-forming ionizing emissivities. Star-forming galaxies dominate the ionizing emissivity at $z > 5$. The observationally-motivated \nion\ is statistically consistent with the observed redshift 4 and 5 points, in marginal statistical tension with the $z\sim6$ data point, and suggests that star-forming galaxies began to emit sufficient ionizing photons to shift the IGM ionization state near redshifts of 8. }
\label{fig:nion}
\end{figure}

The LzLCS \betaobs-\fesc\ relationship and the empirical  scaling between \betaobs\ and \muvobs\ can estimate \nion\ at high-redshift. We use the luminosity function from \citet{bouwens15} and integrate from -24 to -16~mag (or approximately 0.01L$^\ast$) for $z = $4, 5, 6, 7, and 8 respectively. We use the luminosity functions from \citet{bouwens15} largely for consistency with the observed \muvobs-\betaobs\ relations (\autoref{eq:beta_bouwens}), but other luminosity functions provide similar results at these \muvobs\ values \citep{livermore17, atek18}. 

First, we follow previous work and assume a constant log($\xi_{\rm ion}) = 25.27$ \citep{bouwens16}  measured for a population of galaxies at $z\sim 4-5$. This $\xi_{\rm ion}$ is broadly consistent with values used in theoretical studies \citep{robertson13, robertson15, naidu19}. In this base case,  we assume a constant \fesc\ for all galaxies of 10\% (dark gold line in \autoref{fig:nion}) and 20\% (light gold line in \autoref{fig:nion}), again similar to previous work \citep{robertson13, robertson15}. The gray line in \autoref{fig:nion} estimates the \nion\ needed to balance hydrogen ionization and recombinations in the IGM at a given redshift \citep{madau99, ouchi}, assuming a clumping factor that evolves with redshift \citep{shull12}. If \nion\ is greater than the gray line, ionizations occur more frequently than recombinations and the ionization state of the IGM increases. When \nion\ is less than this gray line, recombinations occur more frequently than photoionizations, and the ionization state of the IGM becomes more neutral (or does not change if the IGM is already neutral). Future work will refine these prescriptions and use semi-analytic models to solve the differential equations required to quantify the impact of the LzLCS prescriptions on the IGM reionization history (Trebitsch in preparation). 

The gold lines roughly intersect with the gray line between $z \sim7-8$. This suggests that star-forming galaxies with constant properties produce enough ionizing photons to increase the IGM ionization fraction near redshifts of 8 \citep{ouchi, robertson13, robertson15}. The ionizing emissivity of star-forming galaxies continues to rise at lower redshifts as the number density of galaxies increases. 

The LzLCS and high-redshift data suggest that \fesc\ is not constant, but rather varies with \betaobs, \ms, and \muvobs. Does the LzLCS \betaobs-\fesc\ relation and high-redshift \betaobs\  observations suggest a similar production of ionizing photons as these constant assumptions? Using the \citet{bouwens14} \betaobs-\muvobs\ relations (\autoref{eq:beta_bouwens}), we test two LzLCS relations: (1) the LzLCS \betaobs-\fesc\ (\autoref{eq:bouwens_fesc}) with a constant log(\xiion)~=~25.27 and (2) the LzLCS \betaobs-\xiion$\times$\fesc\ relation (\autoref{eq:beta_fesc_xi}). We integrate the product of these relations and the luminosity functions from -23 to -16~mag. These two predicted \nion\ values track each other within 0.2~dex (the constant \xiion\ has lower \nion), and we only include the relation using \xiion$\times$\fesc\ in \autoref{fig:nion} for clarity (the light blue points in \autoref{fig:nion}).

The empirically-motivated \nion\ values exceeds the value to balance recombinations between $z\sim 7-8$. Near these redshifts, star-forming galaxies produce sufficient ionizing photons to increase the IGM ionization fraction. While we currently do not have statistical populations of \betaobs\ at $z>8$, recent work suggests that the normalization of the luminosity function continues to decline at higher redshifts \citep{oesch18, bouwens21, Finkelstein22, bouwens22}. Unless $z > 8$ galaxies are significantly bluer than $z\sim7$ galaxies, the $z > 8$ galaxy populations will emit an insufficient number of ionizing photons to exceed recombinations in the IGM.

Between redshifts of 3 and 8, \nion\ from star-forming galaxies varies by less than a factor of 2. This constant \nion\ occurs even though $\rho_{\rm UV}$ increases by nearly a factor of 10 from $z\sim 8$ to $z\sim3$. The dwindling \fesc\ at lower redshift  (\autoref{fig:fesc_muv})  balances this increasing number of galaxies to keep \nion\ relatively constant. Meanwhile, \nion\ from AGN increases at lower redshifts (red line), such that star-forming galaxies and AGN have similar \nion\ at $z < 5$. At $z\sim3-4$ star-forming galaxies and AGN contribute nearly equally to the \ion{H}{i} ionizing photon budget \citep{steidel18, dayal20, trebitsch21}. The total (AGN plus star-forming) \nion\ (dark blue circles) match both the low-redshift observations of the ionizing emissivity (gray squares) and the \nion\ value required for ionizations to exceed recombinations in the IGM (gray line) without overproducing ionizing photons at lower redshift. The observationally-motivated \betaobs\ prescription of \fesc\ provides an empirically-motivated estimation of the ionizing emissivity of high-redshift star-forming galaxies and suggests that star-forming galaxies produced sufficient ionizing photons to increase the IGM ionization fraction $z \sim 7-8$. 

Inferring \nion\ at high-redshifts using the LzLCS \betaobs-\fesc\ relations relies on a few critical assumptions extending from $z\sim0.3$ to $z\sim6-9$. First, the dust extinction law sensitively impacts the shape of the FUV continuum, which connects \betaobs\ to the absorption of ionizing photons at 912~\AA\ (see Appendix~\ref{ebv}). Dust properties may evolve strongly over time as different elements -- specifically Fe, C, Si, and  Mg -- have different formation mechanisms and are produced on different timescales. The strong correspondence between \betaobs\ and both \muvobs\ and \ms\ suggests that there are similarities between galaxy properties and dust properties at high- and low-redshifts, but there could be significant differences. For instance, \citet{Saldana} found \fesc\ to be a median factor of 1.08 higher using the \citet{reddy16} extinction law than the SMC law. This factor would slightly impact the ionizing emissivity of star-forming galaxies, but would be insufficient to qualitatively change their ability to reionize the early IGM. If the high-redshift dust extinction law is observed to significantly deviate from the \citet{reddy16} law, Appendix~\ref{ebv} and \autoref{beta} illustrate how the \fesc\ relations can be updated in the future to include these reformulated high-redshift dust laws. Second, observing the range of \betaint\ of high-redshift samples will determine whether \betaint\ strongly deviates from the LzLCS distribution. \autoref{eq:beta_ebv} demonstrates that \betaint\ is required to connect \ebv\ and \betaobs; even though \betaint\ has a narrow range in the LzLCS, it  must also be constrained at high-redshift. Third, the current high-redshift \betaobs-\muv\ relations have appreciable uncertainties. For instance, the slope of the $z\sim7$ \betaobs-\muv\ from \citet{bouwens14} is $0.2\pm0.07$, implying that the relative scaling of the \fesc-\betaobs\ at high-redshift is still modestly uncertain. Large samples from JWST observations will dramatically improve these relations and tighten our understanding of \fesc\ during the epoch of reionization. Finally, using the LzLCS \betaobs\ to determine \fesc\ relies on high- and low-redshift galaxies having similar neutral gas properties. The ionization states of high-redshift galaxies are currently under-constrained and cannot be compared to the LzLCS. Future LzLCS studies will attempt to disentangle additional correlations between \fesc, \betaobs, and the ionization properties of the LzLCS (Jaskot et al. in preparation). As  the JWST makes similar observations at high-redshift, the correlations presented here can be tailored to the discovered conditions at high-redshift.  
 
\section{Summary and Conclusions}
\label{summary}

Here we presented a new analysis of the far-ultraviolet stellar continuum slope ($\beta$) of the Low Redshift Lyman Continuum Survey (LzLCS). We fit 89 galaxies that have \textit{Hubble} Space Telescope LyC observations with stellar population synthesis models to estimate the intrinsic (\betaint) and observed (\betaobs) stellar continuum slope at 1550~\AA. These stellar population fits are critical to determine \betaobs\ at the non-ionizing wavelengths that are typically observed at high-redshifts. Comparing the observed \ovi\ stellar wind profiles to stellar population models demonstrates that the LzLCS is characterized by young stellar populations (\autoref{fig:ovi}). We calculated many properties from the stellar population fits, but focused on \betaobs\ because it is currently commonly observed at high-redshift. 


We then explored trends between \betaobs\ and various properties of the LzLCS. Our main findings are:
\begin{enumerate}
    \item The \betaobs\ scales with the [\ion{O}{iii}]/[\ion{O}{ii}] flux ratio (7.5$\sigma$ significance; \autoref{fig:o32_oh}, left panel), 12+log(O/H) (3.4$\sigma$; \autoref{fig:o32_oh}, right panel), H$\beta$ equivalent width (6.7$\sigma$), stellar mass (5.9$\sigma$; \autoref{fig:beta_ms}), and the observed FUV absolute magnitude (\muvobs; 3.4$\sigma$; \autoref{fig:beta_muv}). In general, this implies that lower metallicity, higher ionization, lower mass, and fainter galaxies have bluer FUV continuum slopes. 
    \item \betaobs\ is largely set by the color excess, \ebv\ (\autoref{fig:beta_obs}), and does not scale with intrinsic stellar properties (\autoref{fig:beta_int}). Thus, we suggest that the LzLCS \betaobs\ is largely driven by dust attenuation. 
    \item \betaobs\ scales strongly with the Lyman Continuum escape fraction (\fesc) at the 5.7$\sigma$ significance (\autoref{fig:beta_fesc}). Galaxies with \betaobs~=~-2.11, -2.35, -2.60 have population-averaged \fesc~=5, 10, 20\%, respectively. We provide a scaling relation between \betaobs\ and \fesc\ in \autoref{eq:beta_fesc}. 
    \item \betaobs\ also scales with the product of \fesc~$\times$~\xiion\ (\autoref{fig:beta_fesc_xi} and \autoref{eq:beta_fesc_xi}). 
    \item There is appreciable scatter in the \fesc-\betaobs\ relations that scales strongly with \fescrel\ (see color-coding of the points in \autoref{fig:beta_fesc}). These \betaobs-\fesc\ relations are well-suited to estimate population averages, rather than \fesc\ from individual galaxies. 
\end{enumerate}
Assuming these above findings transfer to high-redshift galaxies, and combining them with previous observations of \betaobs\ at high-redshift, we infer the emission of ionizing photons into the Intergalactic Medium during the epoch of reionization (\autoref{reionization}). Specifically we find that:
\begin{enumerate}
    \item The \betaobs\ of high-redshift and low-redshift galaxies have similar relations between both  \muvobs\ (\autoref{fig:beta_muv}) and stellar mass (\autoref{fig:beta_ms}). Low-mass galaxies are bluer at both high and low-redshift (\autoref{fig:beta_muv}). Faint galaxies likely dominated the emission of ionizing photons  during the epoch of reionization. 
    \item \fesc\ increases for fainter \muvobs\ at higher redshifts (\autoref{fig:fesc_muv}). Only galaxies with \muvobs~$>-16$~mag at $z < 5$ have population-averaged \fesc\ greater than 10\%. Galaxies during the epoch of reionization are estimated to have higher \fesc\ than their lower redshift counterparts (compare the blue and gold curves in \autoref{fig:fesc_muv}). This implies that relatively bright galaxies (\muvobs~$> -18$~mag) have \fesc\ population averages greater than 10\%. 
    \item \betaobs\ (and \fesc) likely evolves with time and galaxy mass as galaxies synthesize and retain metals and dust. 
\end{enumerate}
We combine the  LzLCS \betaobs-\fesc\ relation, the high-redshift \betaobs\ observations, and the observed luminosity functions to make empirical estimates of the ionizing photon emissivity at high-redshift (\autoref{fig:nion}). These empirically-motivated prescriptions suggest that galaxies near $z \sim 7-8$ first emitted a sufficient number of ionizing photons to increase the IGM ionization fraction. \nion\ of star-forming galaxies flattens at lower redshifts and varies by less than a factor of 2 from $z =8$ to $z=4$. As such, $z \sim 4$ star-forming galaxies and AGN have similar \nion. The star-forming plus AGN emissivities are consistent with IGM observations at $z \sim 4-5$.

Soon JWST will compile larger and more robust samples of faint galaxies at $z>5$. The empirical relations provided here will reveal whether these galaxy populations produced sufficient ionizing photons to reionize the IGM.  Jaskot et al. (in preparation) will explore whether there exists strong secondary correlations between \fesc\ and parameters such as \betaobs, O$_{32}$, H$\beta$ equivalent width, and stellar mass. These new relations will take advantage of the upcoming spectroscopic capabilities of JWST to provide even more robust constraints on the ionizing emissivity of high-redshift galaxies and their impact on cosmic reionization.

\section*{Acknowledgements}
We acknowledge that the location where this work took place, the University of Texas at Austin, sits on stolen indigenous land. The Tonkawa live in central Texas and the Comanche and Apache moved through this area. We pay our respects to all the American Indian and Indigenous Peoples and communities who have been or have become a part of these lands and territories in Texas, on this piece of Turtle Island.

Support for this work was provided by NASA through grant number HST-GO-15626 from the Space Telescope Science Institute. This research is based on observations made with the NASA/ESA Hubble Space Telescope obtained from the Space Telescope Science Institute, which is operated by the Association of Universities for Research in Astronomy, Inc., under NASA contract NAS 5–26555. These observations are associated with program(s)  13744, 14635, 15341, 15626, 15639, and 15941. STScI is operated by the Association of Universities for Research in Astronomy, Inc. under NASA contract NAS 5-26555. ASL acknowledge support from Swiss National Science Foundation. RA acknowledges support from ANID Fondecyt Regular 1202007. MH is fellow of the Knut and Alice Wallenberg Foundation.


\bsp	
\label{lastpage}
\section*{Data Availability}

All data in this paper are made possible through previous publications. 

\appendix

\section{IMPACT OF THE EXTINCTION LAW}
\label{ebv} 

\begin{figure*}
\includegraphics[width = \textwidth]{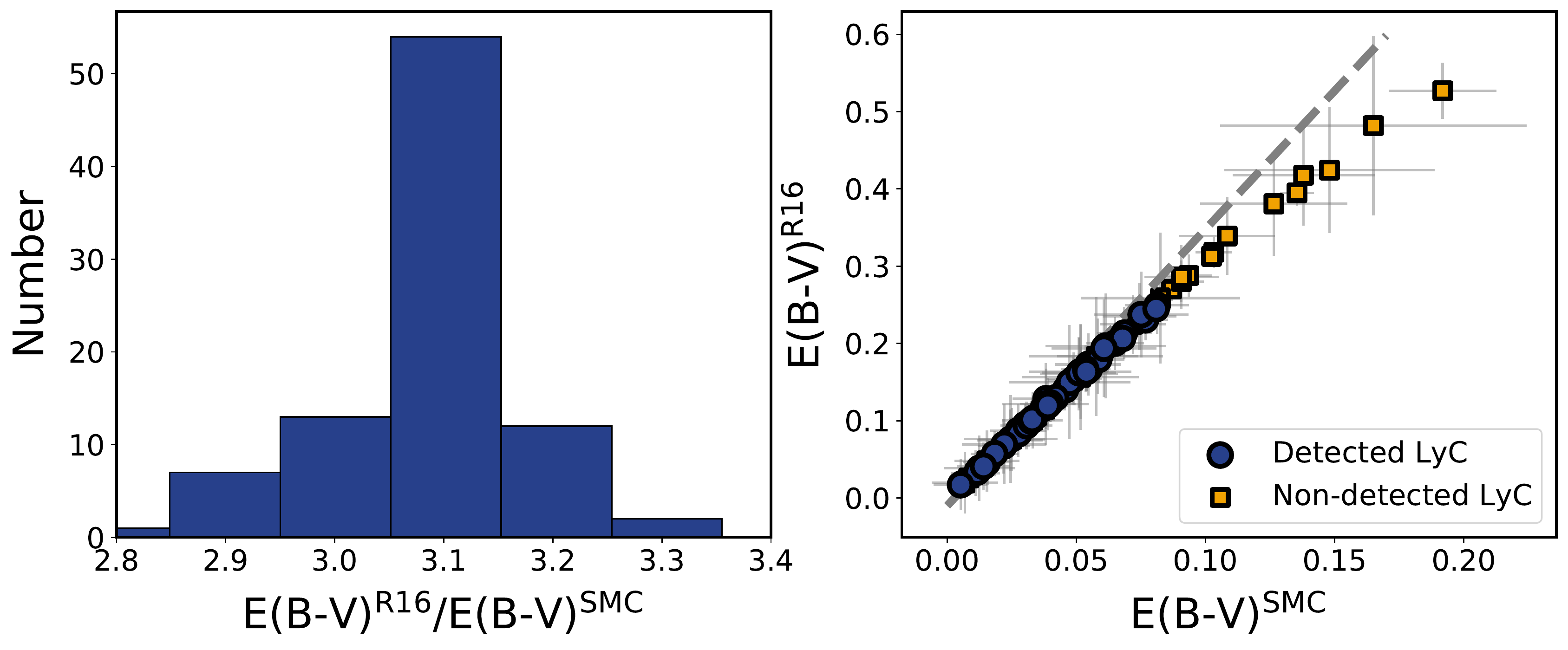}
\caption{The impact of the assumed extinction law on the fitted attenuation parameter. \textbf{Left Panel: } Histogram of the ratio of the attenuation parameters derived using the \citet{reddy16} (R16) and the SMC laws. This ratio peaks strongly at a value of 3.1. \textbf{Right Panel: } The correlation of the attenuation parameter for the R16 and SMC extinction laws for the LyC detected (blue circles) and non-detected (gold squares) samples. Overplotted as a gray dashed line is an analytical fit to this scaling (\autoref{eq:ebv_relation}). }
\label{fig:ebv_comp}
\end{figure*}

The fitted E(B-V) has one of the most statistically significant correlations with \fesc\ \citep{Saldana}. E(B-V) also  strongly determines the \betaobs\ of the LzLCS (see \autoref{beta}), but the E(B-V) parameter  depends on the assumed extinction law. As such an important diagnostic quantity, here we discuss the assumptions underpinning the chosen extinction law and its impact on the inferred spectral shape of the stellar continuum models and the E(B-V) parameter.

In \citet{Saldana}, we tested the impact of the assumed extinction law. We found that varying the extinction law from the \citet{reddy16} to an extrapolation of the SMC law \citep{prevot} did not impact the model parameters that set the intrinsic shape: age and metallicity. Therefore, the intrinsic stellar population properties do not appreciably change with different extinction laws. However, the absolute values of the individual fitted $X_j$ values do change because the SMC extinction law is both steeper and has larger $k(\lambda)$ values than the \citet{reddy16} law. This is likely because the FUV stellar spectral features (e.g. \autoref{fig:ovi}) constrain the intrinsic properties of the stellar population. This sets the intrinsic shape of the stellar continuum, \betaint, and \ebv\ is varied, using the assumed reddening law, to reconcile \betaint\ to  \betaobs\ (see \autoref{beta}).

Comparing the different extinction laws provides a prescription to convert \betaobs\ into E(B-V) for different extinction laws. The left panel of \autoref{fig:ebv_comp} shows the ratio of the \ebv\ assuming a \citet{reddy16} (E(B-V)$^{\rm R16}$) to the \ebv\ assuming an extrapolation of the SMC law (E(B-V)$^{\rm SMC}$). The ratio of the two \ebv\ values peaks strongly with a median value of 3.09 and a standard deviation of 0.09, much smaller than the median 0.7~mag uncertainty on the measured ratio. The right panel of \autoref{fig:ebv_comp} shows the strong correlation between the two \ebv\ values. 

The \ebv\ differences derived with different reddening laws can be estimated analytically. First, we approximate the ratio of the observed flux at two wavelengths (F($\lambda^1$) and F($\lambda^1$)) as the ratio of the intrinsic flux times an unknown attenuation factor as 
\begin{equation}
    \frac{F(\lambda^1)}{F(\lambda^2)} = \frac{F_{i}(\lambda^1)}{F_{i}(\lambda^2)} e^{-E(B-V) \left[k(\lambda^1) - k(\lambda^2)\right]} = R e^{-E(B-V) \Delta k}, \label{eq:ebv_rat}
\end{equation}
where we have defined $R$ as the intrinsic flux ratio at wavelengths 1 and 2, and $\Delta k = k(\lambda^1)-k(\lambda^1)$ is the difference of the extinction law at two wavelengths ($\lambda^1$ and $\lambda^2$). Regardless of the assumed law, the observed flux densities must be equal at the two wavelengths.  Numerically, this means that \begin{equation}
     \frac{F^{\rm R16}(\lambda^1)}{F^{\rm R16}(\lambda^2)}  =\frac{F^{\rm  SMC}(\lambda^1)}{F^{\rm   SMC}(\lambda^2)} .
\end{equation}
Substituting in \autoref{eq:ebv_rat} then suggests that E(B-V)$^{\rm R16}$  is  related to E(B-V)$^{\rm SMC}$ as 
\begin{equation}
    E(B-V)^{\rm R16} = \frac{\Delta k^{\rm SMC}}{\Delta k^{\rm R16}} E(B-V)^{\rm SMC} + \frac{1}{\Delta k^{\rm R16}}\ln{\frac{R^{\rm R16}}{R^{\rm SMC}}}\label{eq:ebv_conversion}
\end{equation}
If we use 950~\AA\ as wavelength 1 and 1200~\AA\ as wavelength 2 (typical wavelengths used to fit the models of the LzLCS), we find that $\Delta k^{R16} =2.1$ and $\Delta k^{\rm SMC} = 6.7$ such that the E(B-V) of the two laws is related as: 
\begin{equation}
    E(B-V)^{R16} = 3.14 \times E(B-V)^{\rm SMC} + 0.47 \ln{\frac{R^{\rm R16}}{R^{\rm SMC}}}\label{eq:ebv_relation}
\end{equation}
The slope of this line matches the median 3.09 found from the ratio of the two attenuation parameters (left panel of \autoref{fig:ebv_comp}). If we fix the slope of the line in \autoref{fig:ebv_comp} to the theoretical ratio of $\Delta k^{\rm SMC}$/$\Delta k^{\rm R16} = 3.14$, we find that $0.47 \ln\left({R^{\rm R16}/R^{\rm SMC}}\right) = 0.004$,  implying that assuming an SMC law leads to an intrinsic FUV flux that is 1.01 times that of the \citet{reddy16} law. Thus, a good approximation for the conversion of the \ebv\ determined in the 950-1200~\AA\ region with the \citet{reddy16} law to the SMC law is a constant factor of 3.1. This exercise illustrates how the individual reddening laws can be converted between each other, depending on the $\Delta k$ values of the individual extinction laws at their specific wavelengths. While we use the \citet{reddy16} law for the paper, \autoref{eq:ebv_conversion} can be used to convert E(B-V) to other dust extinction laws and derive the respective \betaobs\ values.

\bibliographystyle{mnras}
\bibliography{stars} 

\end{document}